\documentclass[epj]{svjour}
\usepackage{graphicx}
\usepackage{bm}
\usepackage{amssymb}
\usepackage{amsmath}
\usepackage{dsfont}
\begin{document}
\title{A circle swimmer at low Reynolds number}
\author{Rodrigo Ledesma-Aguilar\inst{1} \and Hartmut L\"owen\inst{2} \and Julia M. Yeomans\inst{1}% etc
}                     % Do not remove
\institute{The Rudolf Peierls Centre for Theoretical Physics, University of Oxford, 1 Keble Road, Oxford OX1 3NP, United Kingdom  \and Institut f\"ur Theoretische Physik II: Weiche Materie, Heinrich-Heine-Universit\"at D\"usseldorf,
Universit\"atsstra{\ss}e 1, D-40225 D\"usseldorf, Germany}
\date{Received: date / Revised version: date}
% The correct dates will be entered by Springer
%
\abstract{
Swimming in circles occurs in a variety of situations at low Reynolds number.
Here we propose a simple model for a swimmer that undergoes circular motion,
generalising the model of a linear swimmer proposed by Najafi and Golestanian (Phys. Rev. E 69, 062901 (2004)). 
Our model consists of three solid spheres arranged in a triangular configuration, joined by two links of time-dependent length.
For small strokes, we discuss the motion of the swimmer as a function of the separation angle between its links.   
We find that swimmers describe either clockwise or anticlockwise circular  motion depending on the tilting angle in a non-trivial manner.    
The symmetry of the swimmer leads to a quadrupolar decay of the far flow field. 
We discuss the potential extensions and experimental realisation of our model.
 % end of PACS codes
} %end of abstract
\maketitle
\section{Introduction}
\label{intro}
The physics of microswimmers is a rapidly advancing field,
for recent reviews see \cite{ref1,ref2,Cates_2012,Romanczuk2012}.
One of the simplest models of a microswimmer
was proposed by Najafi and Golestanian~\cite{Najafi_2004} in 2004:
it consists of three aligned spheres that 
are linked by rigid rods whose lengths change in time 
between two values. Moving with a periodic motion
which breaks the time-reversal symmetry, this simple swimmer experiences 
a net propulsion along the rod orientation. A number of other swimmer
models have been proposed subsequently~\cite{Avron-NewJPhys-2005,Dreyfus-EPJB-2005,Earl-JChemPhys-2007,Golestanian-EPJE-2008,Golestanian-EJPJE-2008,Alexander-ErophysLett-2008,Lauga-PhysRevE-2008,Leshansky-NewJPhys-2009,Elfring-PhysRevLett-2009}, most of which lead to propulsion along a linear trajectory.

However, there are  many examples in nature showing circle swimming rather than swimming
along a straight line. On a planar substrate, certain
bacteria \cite{ref4,ref5,ref6,ref7,ref8,ref9} and spermatozoa \cite{ref10,ref11,ref12}
swim in circles. Moreover, spherical
camphors have been shown to exhibit circular swimming when confined to an
interface \cite{ref13}. Recently, catalytically or thermally driven colloidal particles
with an asymmetric shape have been prepared \cite{Volpe} also resulting in circular motion
on  a substrate \cite{Hagen_thesis}.    

Even though circle-swimming is frequent
in the presence of solid surfaces, curved trajectories are also very common in the bulk~\cite{Crenshaw-AmerZool-1996}.  
This can be attributed to asymmetries in the swimming stroke that can result in both translational and rotational modes of 
motion~\cite{Purcell}.  Therefore, while linear swimming occurs for highly symmetric swimmers, more generally swimming can occur along curved or circular trajectories.

The modelling of circle swimmers is much less advanced than that of their linearly moving counterparts. For instance,  Dunstan {\it et al.}~\cite{Dunstan}  considered a swimmer model of two spheres with different radii, which is a linear
swimmer in a bulk fluid but yields circle swimming close to a solid surface.  Shum {\it et al.}~\cite{Shum-ProcRoySocA-2010} have implemented a detailed 
model of a flagellate that exhibits circle swimming at surfaces.  
For circle swimmers in the bulk, minimal rotor models~\cite{Dreyfus-EPJB-2005,Earl-JChemPhys-2007,Leoni-EurophysLett-2010,Fily-SoftMatter-2012} or 
very coarse-grained driven Brownian particle models~\cite{Teeffelen_2008,Teeffelen_2009,Hagen_2011_JPCM} have been proposed.
For the latter, the particles proceed with both an effective translational and angular
propagation velocity and experience additional Brownian fluctuations. The deterministic
(noise-free) trajectory in two dimensions is a closed circle. However, a more detailed model
which resolves the hydrodynamic details of the swimming strokes is missing.

In this paper we close this gap and generalise the linear model of Najafi and Golestanian
\cite{Najafi_2004} to a simple circle swimmer. In order to do so, we consider three spheres joined
by two links which are tilted relative to each other and perform the stroke as in the Najafi-Golestanian model.
By using both analytical and numerical methods, we show that the resulting motion
is a closed circle which depends on the swimmer angle, $\beta$, which characterises the separation between the links.  
We  focus on the resulting radius of the trajectories, $R_t$, as a function of $\beta$ for 
strokes of small amplitude.  Interestingly, we find that swimming occurs predominantly along one direction
of the circular trajectory (anticlockwise for the specific configuration studied here), except for a small  
range of angles, where the swimmer reverses the sense of rotation. 
Such a behaviour is robust to changes in the model parameters. 
We further analyse the velocity field produced by this simple circle swimmer
and find a marked inverse-power decay at large distances.   We recover 
the expected quadrupolar far-field behaviour~\cite{Earl-JChemPhys-2007}, with the magnitude of the velocity field decaying as the
inverse cube of the distance from the swimmer.  This is a consequence of the symmetry of the swimming stroke, which is invariant under combined time-reversal 
and parity transformations~\cite{Pooley-PhysRevLett-2007,scattering1}.   For asymmetric swimming strokes, which do not posses the time-reversal and parity symmetry, we recover 
a decay consistent with a dipolar velocity field, as expected. 
  
Our simple model can serve as a starting point for further analytical and numerical studies.
These can include single circle swimmers in confinement \cite{Teeffelen_2008,Teeffelen_2009}
and shear flow \cite{Hagen_2011_PRE}
as well as the scattering~\cite{scattering1,scattering2} and synchronization
\cite{synchro1,synchro2,synchro3} of two circle swimmers and the (still unknown) collective properties of many 
circle swimmers, for example swarming and vortex formation.

The rest of the paper is organised as follows: in Sec.~\ref{sec:Model}, we describe and define the model. A
discussion of the geometry of the resulting trajectory is performed in Sec.~\ref{sec:Trajectories} 
while the velocity fields in the surrounding fluid are discussed in Sec.~\ref{sec:VFields}. Finally
we conclude in Sec.~\ref{sec:Conclusions}.

\section{Model}
\label{sec:Model}

\begin{figure}[t!]
\centering
\includegraphics[width=0.25\textwidth]{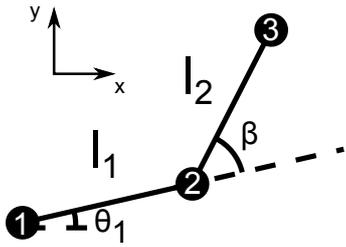} 
\caption{Model of a circle swimmer. Three spherical beads of identical radius $R$ are joined by two links of variable length, 
$l_1$ and $l_2$.  The angle between the links, $\beta \neq 0$, allows the swimmer to move in a curved trajectory. \label{fig:Diagram}}
\end{figure}

Our model swimmer consists of three  beads connected by two massless links.   These are separated by the swimmer angle, $\beta$, and have 
lengths, $l_1$ and $l_2$, which are known functions of time, and hence determine the swimming stroke.\\ 

The configuration of the swimmer can be characterised by the position vectors of the beads, $\mathbf r_i$, which 
are related by the conditions 
$\mathbf r_{12}  = l_1 \mathbf{\hat t}_1,$
and 
$\mathbf r_{23}  = l_2 \mathbf{\hat t}_2,$
where $\mathbf r_{ij} \equiv \mathbf r_j - \mathbf r_i$ and $\mathbf{\hat t}_i$ is the unit tangent vector
to link $l_i$. 
Since, by symmetry, motion can only occur in the plane defined by 
$\mathbf{\hat t}_1$ and $\mathbf{\hat t}_2$, the orientation 
of the swimmer is determined by a single angle.  Here we choose the polar angle 
associated with $\mathbf{\hat t}_1$, $\theta_1$;  the polar angle associated 
with $\mathbf{\hat t}_2$ hence obeys 
$\theta_2 = \theta_1+\beta.$
The swimming stroke imposes kinematic conditions for the bead velocity vectors, $\mathbf v_i \equiv \dot{\mathbf r}_i$, where the dot indicates differentiation
with respect to time.  For the tangential motion, corresponding to the contraction and extension of the links, 
\begin{equation}\label{eq:tangent1}
(\mathbf v_2 -\mathbf v_1)\cdot \mathbf{\hat t}_1 = \dot l_1,
\end{equation}
and
\begin{equation} \label{eq:tangent2}
(\mathbf v_3 -\mathbf v_2)\cdot \mathbf{\hat t}_2 = \dot l_2,
\end{equation}
while for the angular velocities 
\begin{equation}\label{eq:rotationalv}
\frac{(\mathbf v_2 -\mathbf v_1)\cdot \boldsymbol{\hat n}_1}{l_1}  - \frac{(\mathbf v_3 -\mathbf v_2)\cdot \boldsymbol{\hat n}_2}{l_2} = \dot \beta,
\end{equation}
where $\dot{\mathbf{\hat t}}_i = \dot{\theta}_i \boldsymbol{\hat n}_i $.\\ 

The bead dynamics in the overdamped limit is given by 
\begin{equation}\label{eq:dynamics}
{\mathbf v}_i = \sum_{j=1}^3 \mathbf H_{ij}\cdot \mathbf F_j,
\end{equation}
where the velocity of bead $i$ results from the force acting on each bead, 
$\mathbf F_j$, mediated by the hydrodynamic interaction tensor, $\mathbf H_{ij}$, summed over all beads.  
For a Newtonian fluid of viscosity $\eta$, the hydrodynamic interactions can be described using the Oseen tensor,  
\begin{equation}\label{eq:Oseen}
\mathbf H_{ij} = \left\{
\begin{array}{lcr}
\ \mathds 1 / 6\pi \eta R & \mathrm{if} & i= j, \\
1 / 8\pi \eta r_{ij}\left(\mathds 1  + \mathbf r_{ij}\mathbf r_{ij}/r^2_{ij}\right) & \mathrm{if} & i\neq j,
\end{array}
\right.
\end{equation}
which is valid in the limit $R/r_{ij}\ll1$.  

To complete the model, we impose force-free and torque-free conditions on the swimmer,
\begin{equation}\label{eq:forcefree}
\sum_{i=1}^3 \mathbf F_i = 0,
\end{equation}
and 
\begin{equation}\label{eq:torquefree}
\sum_{i=1}^3 \mathbf r_i \times \mathbf F_i =0.
\end{equation}
  
Eqs.~(\ref{eq:tangent1})-(\ref{eq:torquefree}) constitute a linear system for the bead velocities and forces, whose solution depends on time 
through $l_1(t)$, $l_2(t)$ and $\beta(t)$ - which are prescribed functions of time - and through $\theta_1(t)$, which reflects the dependence
on the particular frame of reference and satisfies $\dot \theta_1 = \left(\mathbf v_2 -\mathbf v_1\right)\cdot \boldsymbol{\hat n}_1/l_1.$
Once $\theta_1(t)$ is known, the evolution of the swimmer configuration and force distribution can be obtained by time integrating the $\mathbf{v}_i$
and $\mathbf F_i$.

\begin{figure}
\centering
 \includegraphics[width=0.3\textwidth]{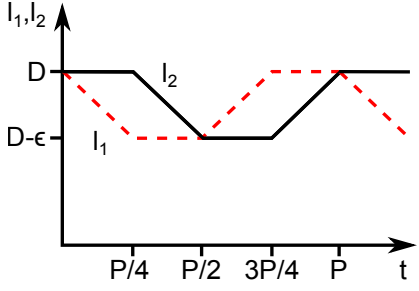}
\caption{Piece-wise swimming stroke.  The links of the swimmer, of maximum extension $D$, contract and expand out of phase by a small length $\epsilon$ (exaggerated in the figure) 
over a period $P$.  
\label{fig:Stroke}}
\end{figure}

Following Refs.~\cite{Najafi_2004} and~\cite{Earl-JChemPhys-2007} we consider a four-step swimming stroke, 
where the links contract and expand alternately at constant velocity $W$ from an initial extension $D$ with a change in length $\epsilon.$
This has the advantage of simplifying the analytics considerably, although other choices of the swimming stroke, consisting for instance in a 
continuous sinusoidal link variation, are also possible.   
The time evolution of the length of the links is chosen to be
\begin{equation}\label{eq:stroke1}
l_1(t) = \left\{
\begin{array}{lcr}
D-Wt & \qquad {\rm if} \qquad& 0 \leq t < P/4, \\
D-\epsilon & {\rm if} & P/4 \leq t < P/2, \\
D-\epsilon+Wt & {\rm if} & P/2 \leq  t < 3P/4, \\
D & {\rm if} & 3P/4 \leq t < P, \\
\end{array}
\right.
\end{equation}
and 
\begin{equation}\label{eq:stroke2}
l_2(t) = \left\{
\begin{array}{lcr}
D & \qquad {\rm if}\qquad &  0 \leq t < P/4, \\\
D-Wt & {\rm if} & P/4 \leq t < P/2,\\
D-\epsilon & {\rm if} & P/2 \leq  t < 3P/4, \\
D-\epsilon +Wt & {\rm if} & 3P/4 \leq t < P,\\
\end{array}
\right.
\end{equation}
and is repeated subsequently over a period $P$.  As shown in Fig.~\ref{fig:Stroke}, the swimming stroke dictated by Eqs.~(\ref{eq:stroke1}) and~(\ref{eq:stroke2}) breaks the 
time-reversal symmetry, and thus results in a net propulsion of the swimmer.  

\section{Swimmer trajectories}
\label{sec:Trajectories}

We treat the problem analytically by first solving the linear system~(\ref{eq:tangent1})-(\ref{eq:torquefree}),
we subsequently expand the $\mathbf{v}_i$ in powers of $R/D$ (up to first order), and $\epsilon/D$ and $\beta$ (up to third order) and restrict ourselves to the case where $\beta$ is constant 
in time.  
We are interested in the trajectory described by the swimmer, which can be characterised by angular and translational displacements 
over one period of the swimming stroke, $\Delta \theta$ and $\Delta \mathbf r$, respectively.    Given that we expect only a small 
angular displacement within each step of the stroke cycle, it is sensible to shift to the frame of reference where $\theta_1(0)=0$ at each step  
and perform an expansion in powers of $\theta_1$.   Keeping only the leading order term in the expansion (constant angular velocity approximation) and adding 
the contribution of each step of the swimming stroke, we obtain, for the angular displacement,
%
%\begin{multline}\label{eq:adisp}
%\Delta \theta \equiv \int_0^T \dot \theta_1 {\rm d} t \approx \left(\frac{R}{D}\right)\left( 2\left(\frac{\epsilon}{D}\right) ^2 +3 \left(\frac{\epsilon}{D}\right)^3\right)\\
%\times\left(\frac{5}{32}\beta+\frac{13}{1152}\beta^2-\frac{83}{864}\beta^3\right)\\
%+\frac{1}{162}\left(\left(\frac{\epsilon}{D}\right) ^2+ \left(\frac{\epsilon}{D}\right)^3\right)\beta^3.
%\end{multline}
\begin{multline}\label{eq:adisp}
\Delta \theta \equiv \int_0^P \dot \theta_1 {\rm d} t \approx \frac{5}{32}\left(\frac{R}{D}\right)\left( 2\left(\frac{\epsilon}{D}\right) ^2 +3 \left(\frac{\epsilon}{D}\right)^3\right)\\
\times\left(\beta-\frac{77}{180}\beta^3\right).
\end{multline}
For the translational displacement vector of the center of mass, $\Delta \mathbf{r} \equiv \int_0^P \frac{1}{3}\sum_{i=1}^{3} \mathbf{v}_i {\rm d} t = \Delta x\boldsymbol{\hat e_x} + \Delta y\boldsymbol{\hat e_y} $, 
we obtain
%
%\begin{multline}\label{eq:tdisp}
%\Delta \mathbf{r} \equiv \int_0^P \mathbf{v}_2 {\rm d} t  \approx R\left\{\left(\frac{7}{12}+\frac{1}{288}\beta\right)\left(\left(\frac{\epsilon}{D}\right)^2+\left(\frac{\epsilon}{D}\right)^3\right)\right.\\
 %\left.-\frac{1}{36}  \left(2 \left(\frac{\epsilon}{D}\right)^2+\left(\frac{\epsilon}{D}\right)^3 \right)\beta\right\}\boldsymbol{\hat e_x}\\
%+R\left\{\left(\frac{47}{144}\right)\left(\left(\frac{\epsilon}{D}\right)^2+\left(\frac{\epsilon}{D}\right)^3\right)\right.\\  
%\left.+\frac{1}{18}  \left(2 \left(\frac{\epsilon}{D}\right)^2+\left(\frac{\epsilon}{D}\right)^3 \right)\right\}\beta\boldsymbol{\hat e_y}.
%\end{multline}
\begin{multline}\label{eq:tdispx}
\Delta x \approx R\left(\left(\frac{\epsilon}{D}\right)^2+\left(\frac{\epsilon}{D}\right)^3\right)\left(\frac{7}{12}-\frac{53}{144}\beta^2\right)\\
-\frac{5R}{4608}\left(\frac{R}{D}\right)\left(2\left(\frac{\epsilon}{D}\right)^2+33\left(\frac{\epsilon}{D}\right)^3\right)\beta^2,\\
\end{multline}
and
\begin{multline}\label{eq:tdispy}
\Delta y \approx R\left(\left(\frac{\epsilon}{D}\right)^2+\left(\frac{\epsilon}{D}\right)^3\right)\left(\frac{7}{24}\beta-\frac{53}{144}\beta^3\right)\\
-\frac{5R}{9216}\left(\frac{R}{D}\right)\left(2\left(\frac{\epsilon}{D}\right)^2+33\left(\frac{\epsilon}{D}\right)^3\right)\beta^3.\\
\end{multline}
Eqs.~(\ref{eq:adisp})-(\ref{eq:tdispy}) reduce to the result reported by Earl {\it et al.}~\cite{Earl-JChemPhys-2007} for vanishing $\beta$, where the swimmer moves along
a linear trajectory with a displacement per swimming stroke
\begin{equation}\label{eq:tdispNG}
\Delta \mathbf{r}  \approx R\left(\frac{7}{12}\right)\left(\left(\frac{\epsilon}{D}\right)^2+\left(\frac{\epsilon}{D}\right)^3\right)\boldsymbol{\hat e_x}.
\end{equation}  
\begin{figure}
\centering
\includegraphics[width=0.45\textwidth]{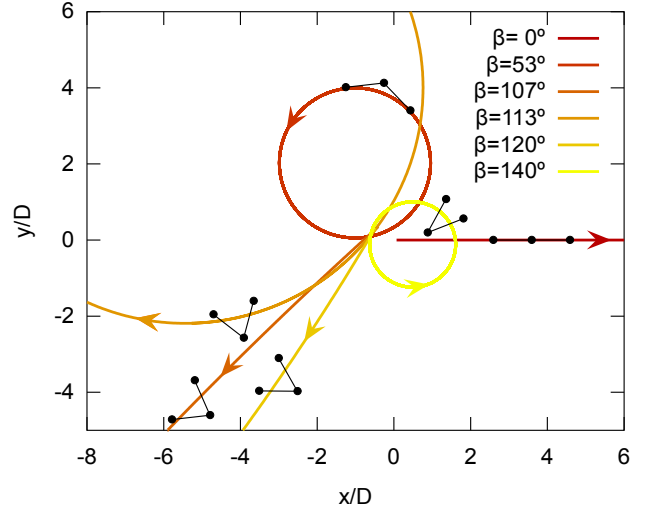}
\caption{Circular trajectories for swimmers at constant { swimmer angle}, $\beta$.  Parameter values are $\epsilon/D=R/D = 10^{-1}$.  \label{fig:trajectories}}
\end{figure}
\begin{figure*}
\centering
\includegraphics[width=0.45\textwidth]{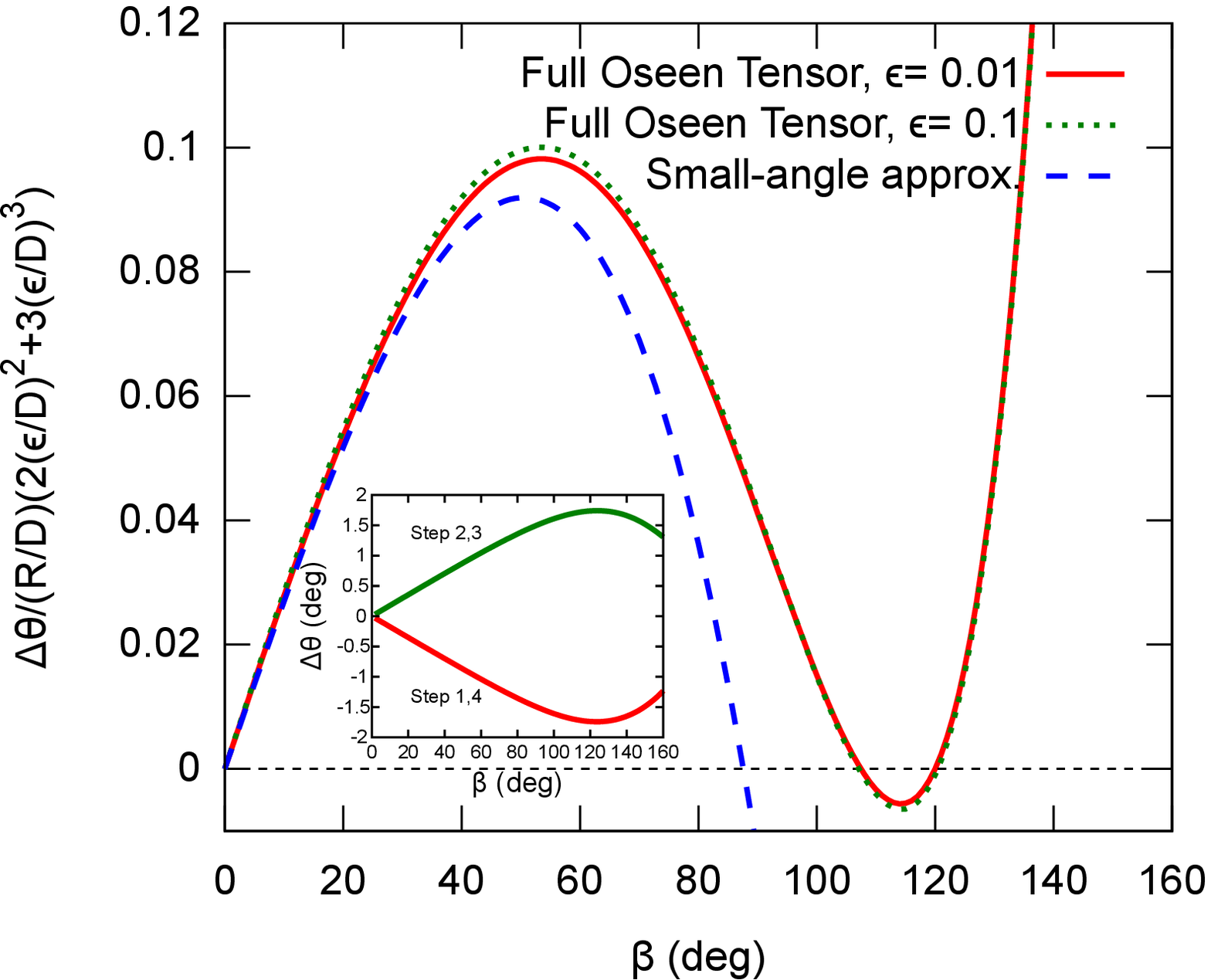} \includegraphics[width=0.45\textwidth]{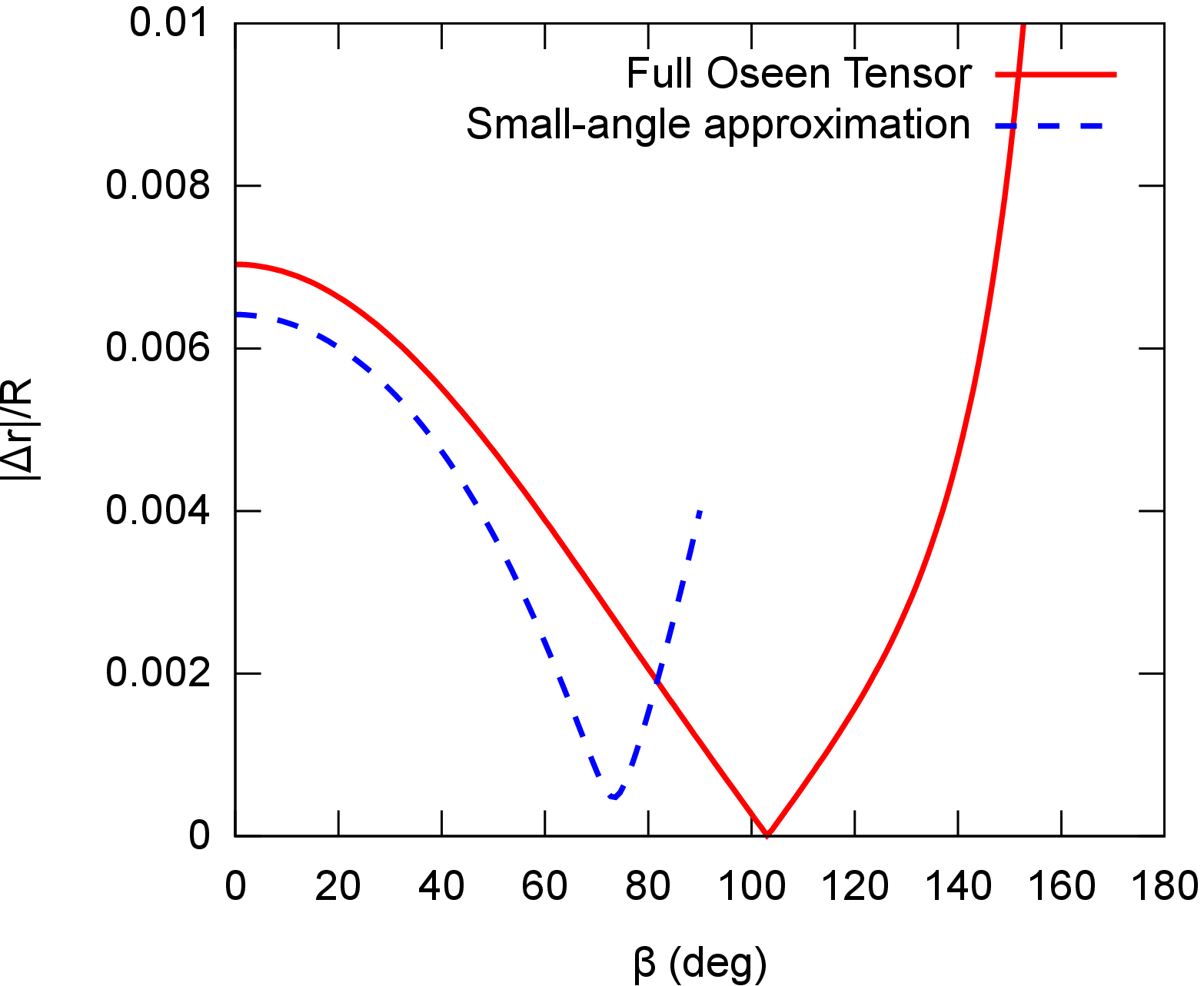}\\
(a)\hspace{6cm}(b)
\caption{(a) Normalised angular displacement as a function of the swimmer angle $\beta$.  Numerical results for the full Oseen-tensor hydrodynamics are carried out 
for $\epsilon=10^{-2}$ (solid line) and $\epsilon=10^{-1}$ (dots).  The dashed curve corresponds to the perturbative result.  Inset:  angular displacement during each step of the swimming
stroke as a function of $\beta$.    (b) Translational displacement as a function of $\beta$ for $\epsilon=10^{-1}$ (solid line).  The dashed curve corresponds
to the perturbative result. \label{fig:AngleRadius}}
\end{figure*}
{ For small, but finite, $\beta$ the trajectory is no longer linear; according to Eqs.~(\ref{eq:adisp})-(\ref{eq:tdispy}) the swimmer undergoes a small positive rotation, and
a translation along the $x$ and $y$ directions, thus describing a curved trajectory.  While $\Delta \theta$ and $\Delta y$ are odd functions of $\beta$, $\Delta x$ is an even
function of the swimmer angle;  this is consistent with the symmetry of the trajectory under the transformation $\beta\rightarrow-\beta$, which corresponds to a reflection about 
the $x$-axis and has the effect reversing the sense of rotation and motion along the $y$ direction while keeping the translation along $x$ unchanged.}  

{ Given that for a given $\beta$ value $\Delta \theta$ and $|\Delta \mathbf{r}|$ are constants, the stroke-averaged trajectories form equilateral chains, becoming regular polygons whenever 
$\Delta \theta = 2\pi/m$ for integer $m$.  Furthermore,  due to the smallness of the angular displacement, the trajectories approach closed circles with curvature}  
\begin{equation}\label{eq:curvature}
\frac{1}{R_t} \approx \frac{\Delta\theta}{|\Delta \mathbf{r}|} \approx \frac{15}{56}\left(2+\left(\frac{\epsilon}{D}\right)-\left(\frac{\epsilon}{D}\right)^2+\left(\frac{\epsilon}{D}\right)^3\right)\frac{\beta}{D}.
\end{equation}

{ In order to verify this assertion}, and to explore the full range of $\beta$, we carry out numerical simulations for the motion of the swimmer, integrating the Oseen-level hydrodynamics over time. Numerical simulations are performed following the algorithm proposed in Ref.~\cite{Earl-JChemPhys-2007}.  At a given timestep, the swimmer shape is first updated 
according to the prescribed swimming stroke, while the position of its centre of mass and orientation are kept constant.  We then use an iterative algorithm to find
the position and orientation that satisfy the force-free and torque-free conditions.  
Swimmer trajectories, with the swimmer superimposed at an arbitrary time, are shown in Fig.~\ref{fig:trajectories}.  { As expected, 
trajectories are very close to circular. Fig.~\ref{fig:AngleRadius} shows plots of $\Delta \theta$ and $|\Delta \mathbf{r}|$ as a function of $\beta$, where 
we also plot the perturbative result, given by Eqs.~(\ref{eq:adisp})-(\ref{eq:tdispy}).   The angular displacement, depicted in Fig.~\ref{fig:AngleRadius}(a), shows a good agreement
with the analytics up to angles as large as $\beta\approx 40^\circ$.  As suggested by Eq.~(\ref{eq:adisp}), results follow the same master curve for different deformations ($\epsilon/D=10^{-1}$ and $\epsilon/D=10^{-2}$) when rescaling by the amplitude $(R/D)(2(\epsilon/D)^2+3(\epsilon/D)^3)$.  This indicates that the shape of the curve is a function of $\beta$ only and, 
consequently, that the location of maxima and minima is independent of both $R/D$ and $\epsilon/D$.  Fig.~\ref{fig:AngleRadius}(b), shows the magnitude of the translational displacement, 
which captures the main qualitative features of the numerical result.   The slight discrepancy at small $\beta$ can be attributed to corrections in $R/D$ and $\epsilon/D$ as discussed in 
Ref.~\cite{Earl-JChemPhys-2007} for linear swimmers.} 

{ While their trajectories are always circular, the sense of rotation of the swimmers changes depending on $\beta$, as shown in Fig.~\ref{fig:AngleRadius}(a).  
%A first group of swimmers,  
%covering the range $0 < \beta \lesssim 107^\circ$,  has positive angular displacements over one period of the 
%swimming stroke.}  Swimmers in this group therefore move in the counter-clockwise direction along their specific trajectory. The second group, lying in the range $107^\circ \lesssim \beta \lesssim 120^\circ$, consists of swimmers that describe a clockwise motion.  Finally, the third group corresponds to $\beta > 120^\circ$,  with the swimmer again moving in the counter-clockwise direction
%along  its trajectory.   
Such a dependence results from the competition between the four steps in the swimming stroke.  During step 1, 
$l_1$ contracts and bead 3 moves to the left, experiencing a drag pointing to
the right (see Fig.~\ref{fig:Diagram}).  As a consequence, there is a torque acting on $l_1$ that causes a negative rotation of the swimmer.  A similar reasoning can be used to conclude that the angular 
displacement must be positive for step 2.    Step 3 can be mapped onto step 2 by performing combined time-reversal and parity transformations, and therefore gives rise to 
the same positive angular displacement.  Similarly, step 4 can be mapped to step 1.  This is verified in the inset of Fig.~\ref{fig:AngleRadius}(a),  where we superimpose the angular 
displacement at each step of the swimming stroke as a function of $\beta$.  Given that for steps 2 and 3 the relative distances between the beads are smaller than
for steps 1 and 4, the angular displacement tends to be larger for the former.  The net displacement is therefore positive for a wide range in $\beta$.  For the range
$107^\circ \lesssim \beta \lesssim 120^\circ$ the numerics show that steps 1 and 4 dominate, causing negative net angular displacements.}  

\begin{figure}[t!]
\centering
\includegraphics[width=0.45\textwidth]{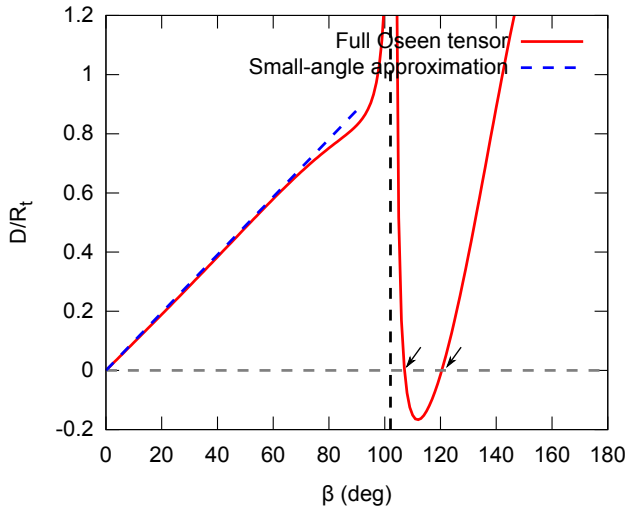}
\caption{Normalised curvature of the swimmer trajectory as a function of the { swimmer angle}, $\beta$.  The solid line corresponds to the 
numerical solution of the Oseen-tensor hydrodynamics while the dashed curve corresponds to the perturbative result.  The arrows indicate the $\beta$ values for which 
swimmers describe a linear motion.  The vertical asymptote indicates a purely rotating swimmer.  Model parameters are $R/D=10^{-1}$ and $\epsilon/D=10^{-1}$.  \label{fig:Radius}}
\end{figure}
%The initial and final steps correspond
%to the contraction of $l_1$ with $l_2$ fully extended, and the expansion of $l_2$ with $l_1$ fully extended, respectively.  Regardless 
%of $\beta$, during these steps the swimmer always undergoes negative angular displacements, {\bf as shown in the inset of Fig.~\ref{fig:AngleRadius}(a).}   
%Conversely, during the second and third steps of the stroke,  $l_2$ first contracts, with $l_1$ already at its minimum value, and subsequently $l_1$ expands, 
%with $l_2$ fully contracted.   This results in positive angular displacements.  
%For $107^\circ \lesssim \beta \lesssim 120^\circ$, the intermediate steps dominate the angular motion of the swimmer, while the first and fourth steps do so 
%for all other values of $\beta$.  

{ Based on their sense of rotation, we can divide swimmers into three main groups.}  The first group, corresponding to positive $\Delta \theta$, is delimited by two vanishing points, located at $\beta=0^\circ$ and $\beta \approx 107^\circ$, with a maximum 
located at $\beta\approx 53^\circ$.  Within the same range of angles, the translational displacement, $|\Delta \mathbf{r}|$, decreases with increasing $\beta$ to a minimum located at  $\beta\approx 103^\circ$ and then increases again.  For the second group of swimmers the angular displacement is negative, as shown in Fig.~\ref{fig:AngleRadius}(a), and also has a non-monotonic behaviour, 
here $\Delta \theta$ decreases for $\beta> 107^\circ$ up to a minimum value, roughly located at $\beta = 113^\circ$, and then increases 
again, vanishing at $\beta\approx 120^\circ$.   In this same range the translational displacement increases monotonically for angles larger than $103^\circ$.
For $\beta>120^\circ$, corresponding to the third group of swimmers, both the angular and translational displacements increase monotonically.   These results are summarised in Fig.~\ref{fig:Radius}, where we depict the reciprocal radius of the swimmer trajectory, $D/R_t$, as a function of the { swimmer angle, $\beta$}, (in units of the rest link length, $D$).  
The sign of $D/R_t$ reflects the sense of motion along the trajectory, being clockwise for $D/R_t<0$ and anti-clockwise if $D/R_t > 0$.  Angles for which swimmers have linear trajectories, corresponding 
to a vanishing angular displacement in Fig.~\ref{fig:AngleRadius}(a), are indicated by arrows.  The purely rotating swimmer, with vanishing $|\Delta\mathbf{r}|$ but finite $\Delta \theta$ is indicated by the asymptote 
located at $\beta\approx 103^\circ$.    

Apart from giving a useful insight into the circle swimming exhibited by the model, Fig.~\ref{fig:Radius} can be used as a read-out to 
choose a particular swimmer, depending on the desired radius of trajectory and sense of motion.  This can be useful given the wide range of 
radii that the trajectories can adopt.    While the range in $\beta$ that results in clockwise motion is rather narrow, one can always obtain the desired 
direction of motion by considering negative values of $\beta$ (mirror image swimmer). 

\section{Velocity fields}
\label{sec:VFields}

Based on the qualitative change in the trajectory with the { swimmer angle}, we expect that the flow field created by the swimmers 
also changes with $\beta$.  This is interesting in terms of the hydrodynamic interactions with surrounding objects, {\it e.g.} passive particles, both small tracers and 
extended objects, and other swimmers.    The average velocity field is calculated as
\begin{equation}
\bar{\mathbf{v}}_i = \frac{1}{T}\int_0^T\sum_j \mathbf{H}_{ij} \cdot \mathbf F_j \mathrm d t,
\end{equation}
where $i$ denotes a point in space with position vector $\mathbf {r}_i$ in the frame of reference of the hydrodynamic centre of the swimmer with the 
links oriented at an angle $\beta/2$ with respect to the $x$-axis.  

\begin{figure*}
\centering
\includegraphics[width=0.4\textwidth]{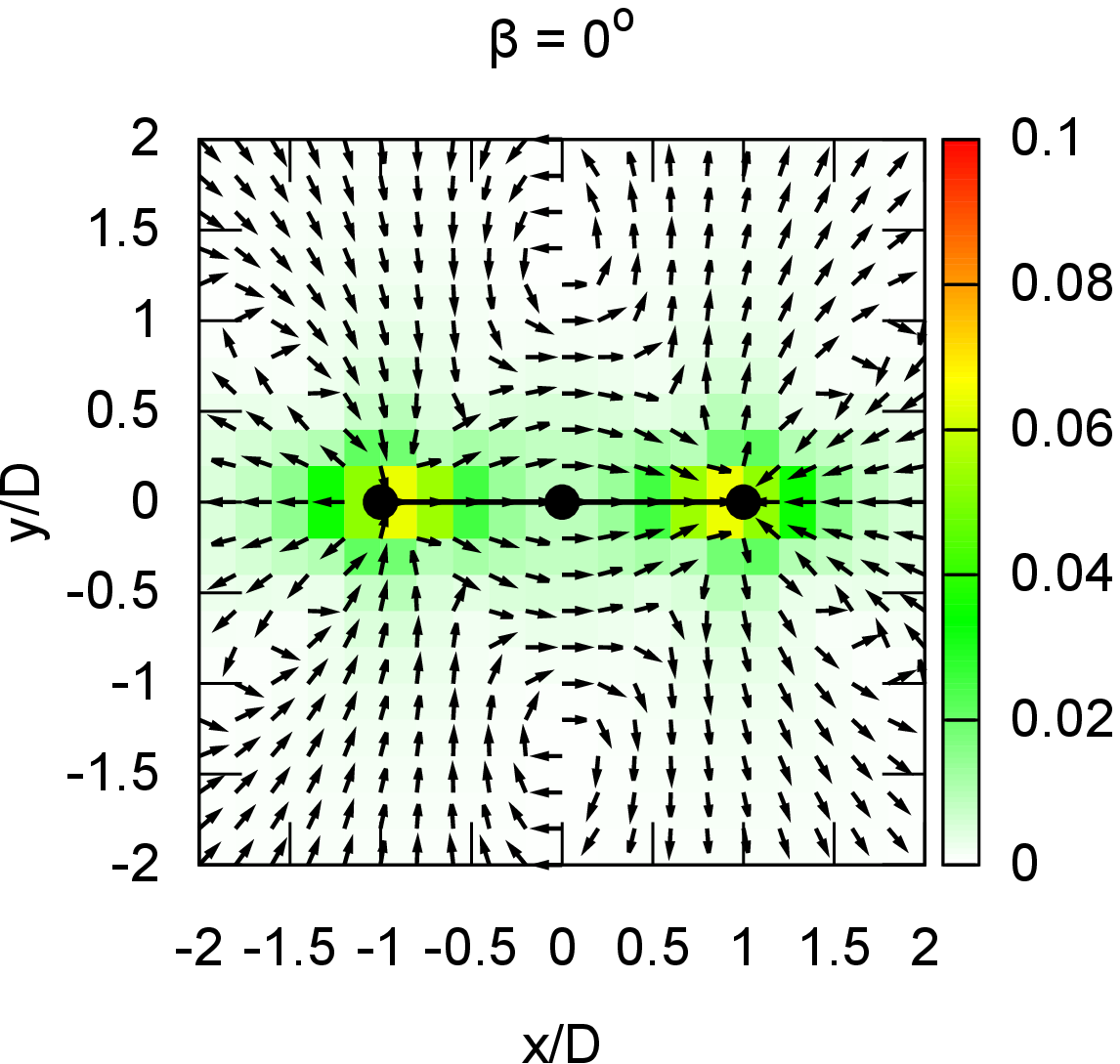}  \includegraphics[width=0.4\textwidth]{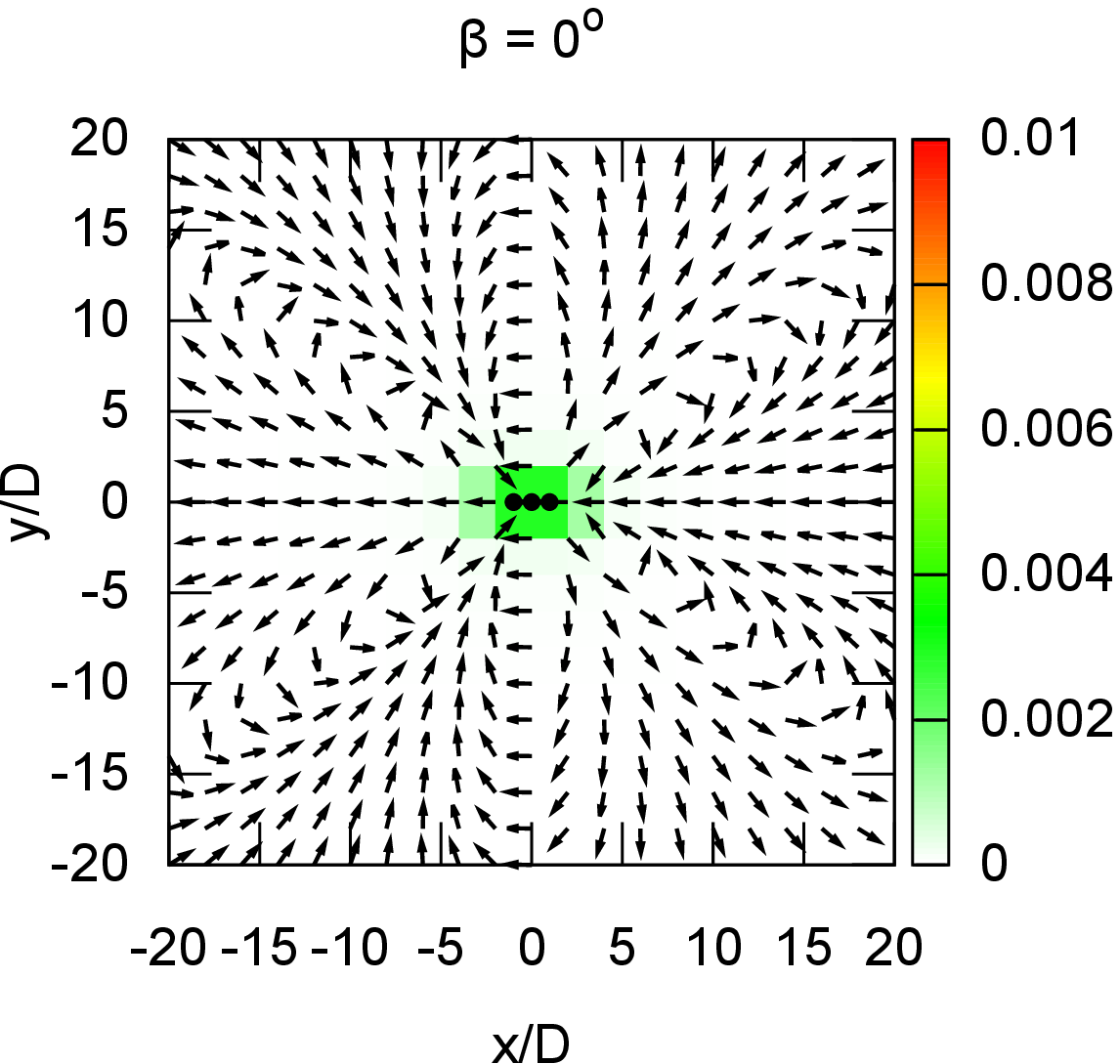}\\
\includegraphics[width=0.4\textwidth]{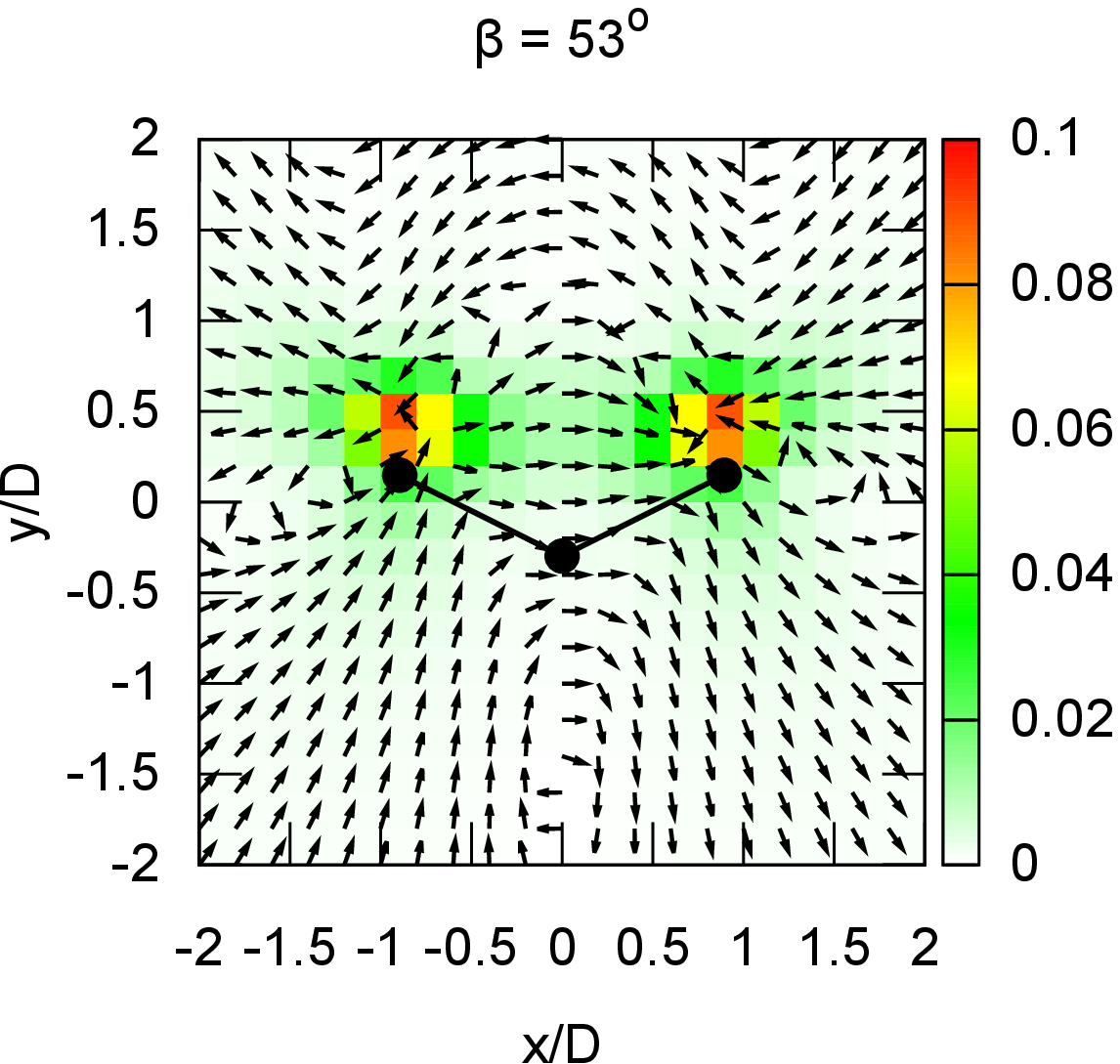}  \includegraphics[width=0.4\textwidth]{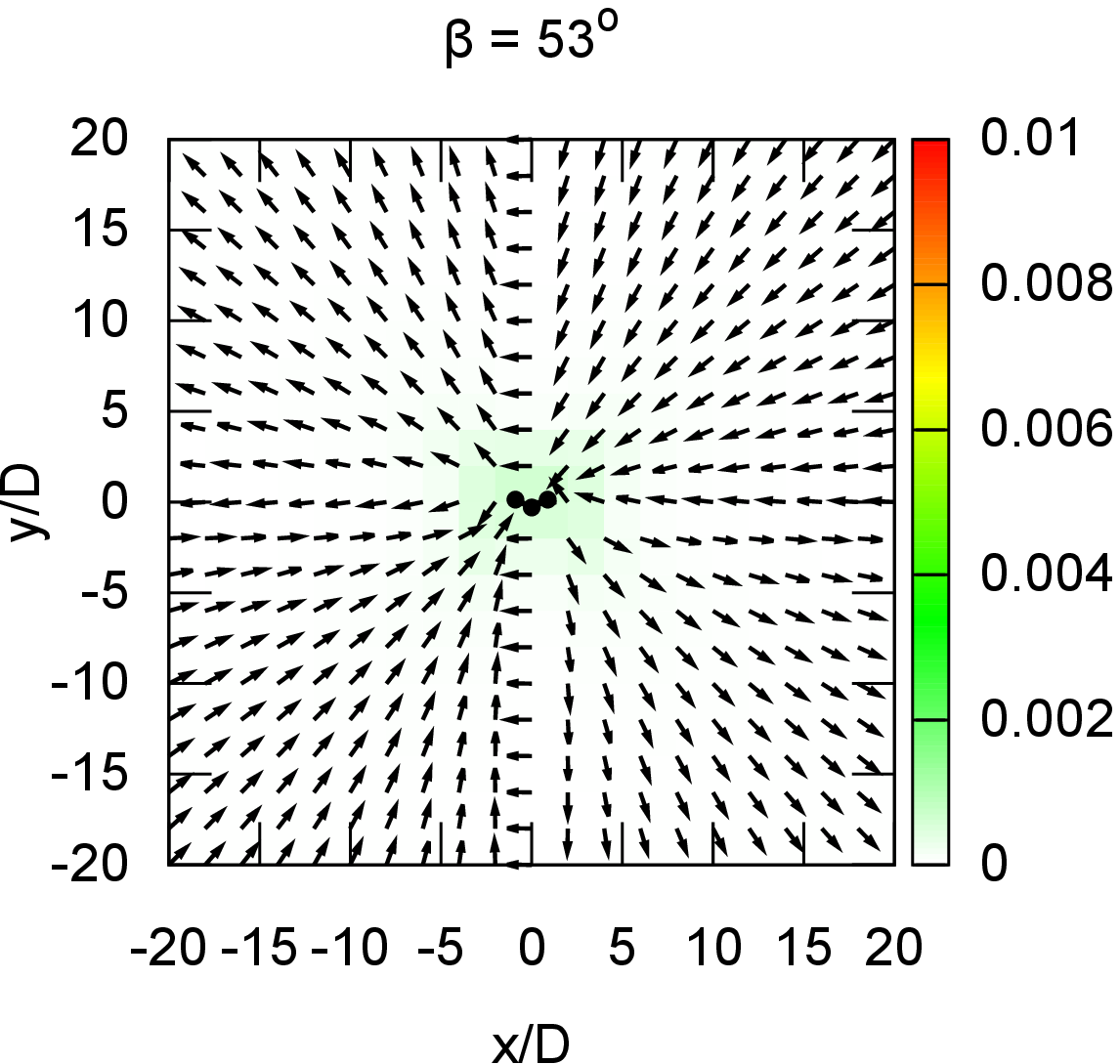}\\
\includegraphics[width=0.4\textwidth]{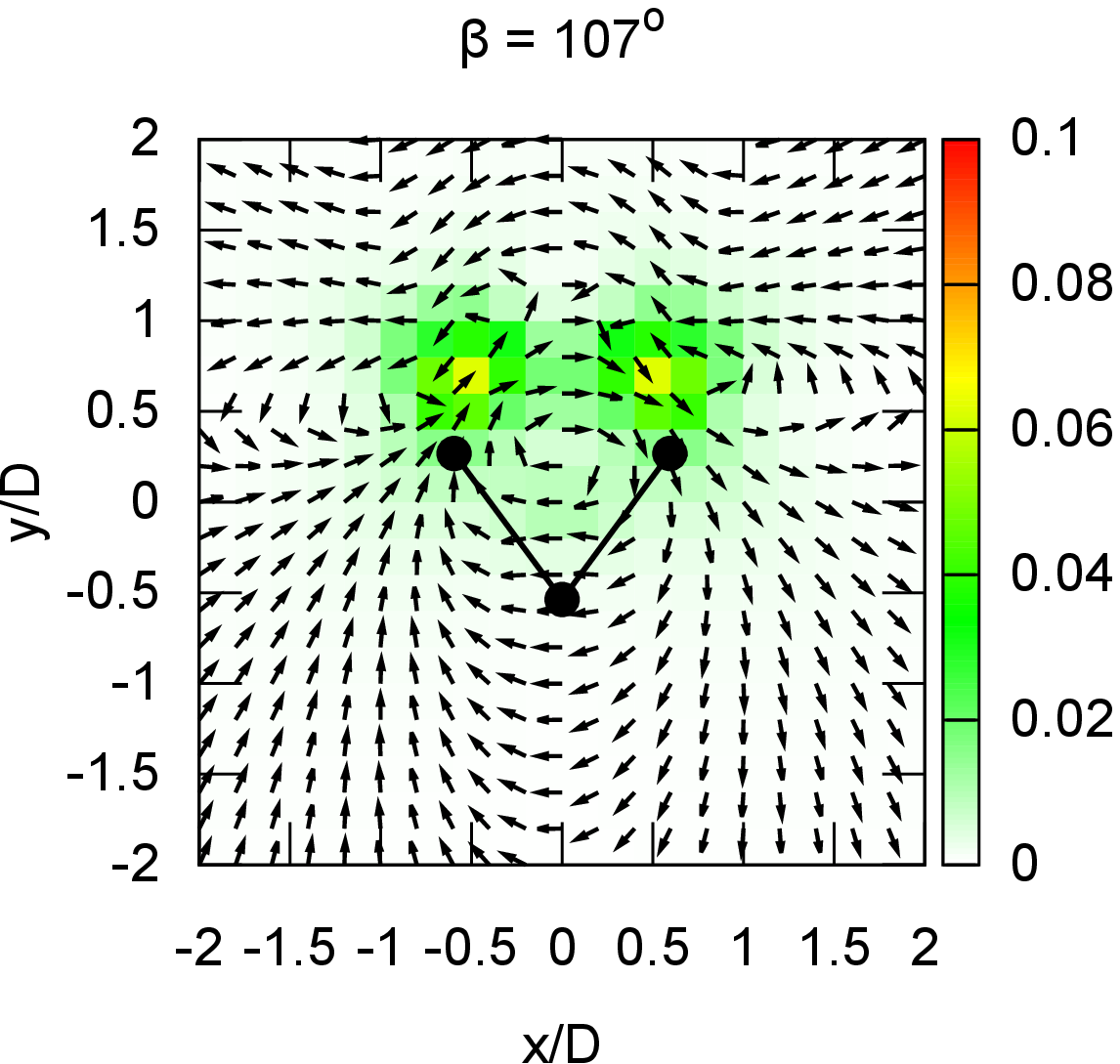}  \includegraphics[width=0.4\textwidth]{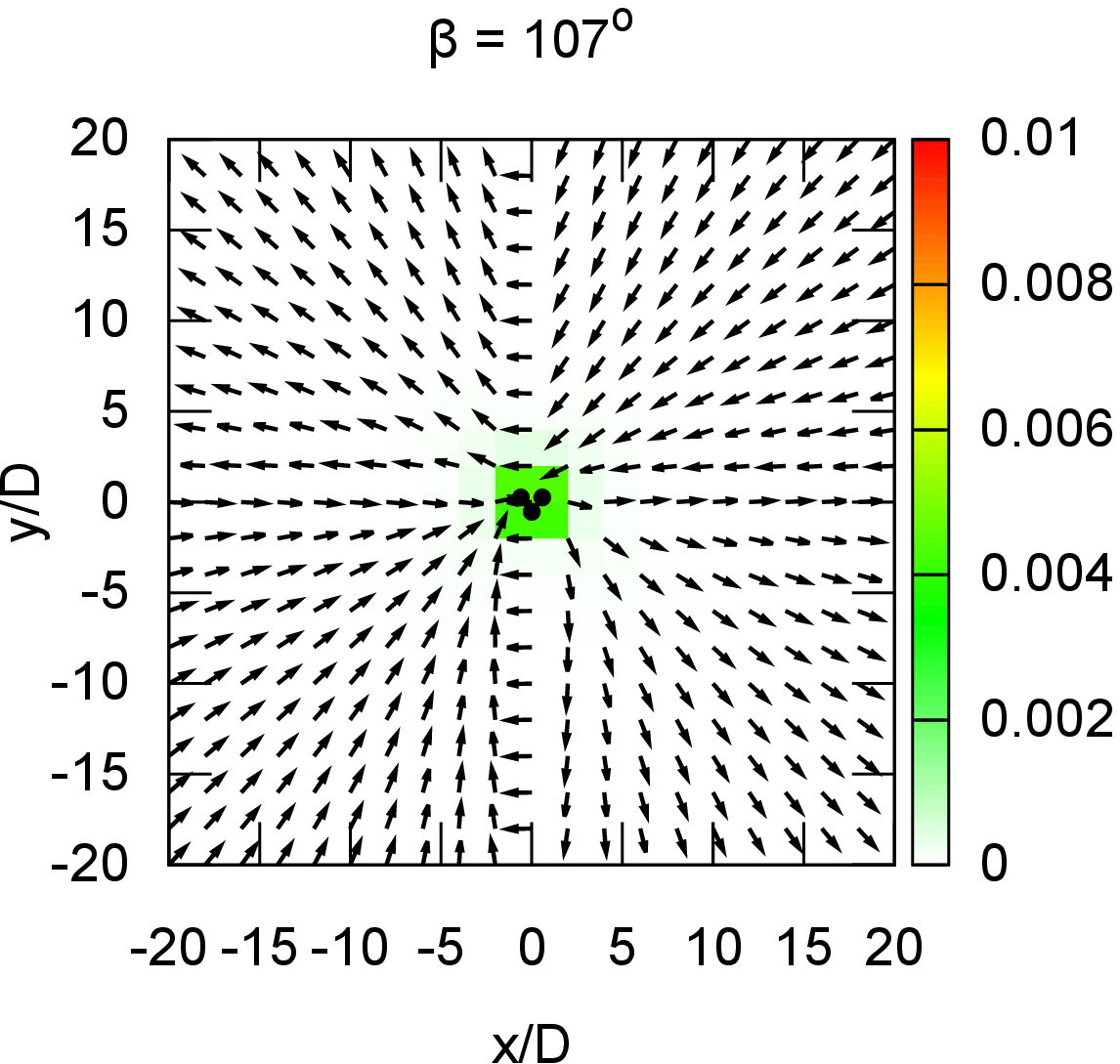}\\
\caption{Average velocity fields near (left) and far (right) from the swimmer for $\beta=0^\circ$, $53^\circ$, and $107^\circ$.  Arrows indicate the 
direction of the velocity field, $\mathbf v_i/|\mathbf v_i|$,while the colour scale indicates its normalised magnitude, $P|\mathbf v_i|/\epsilon$.   \label{fig:velocityfields1}}
\end{figure*}

\begin{figure*}
\centering
\includegraphics[width=0.4\textwidth]{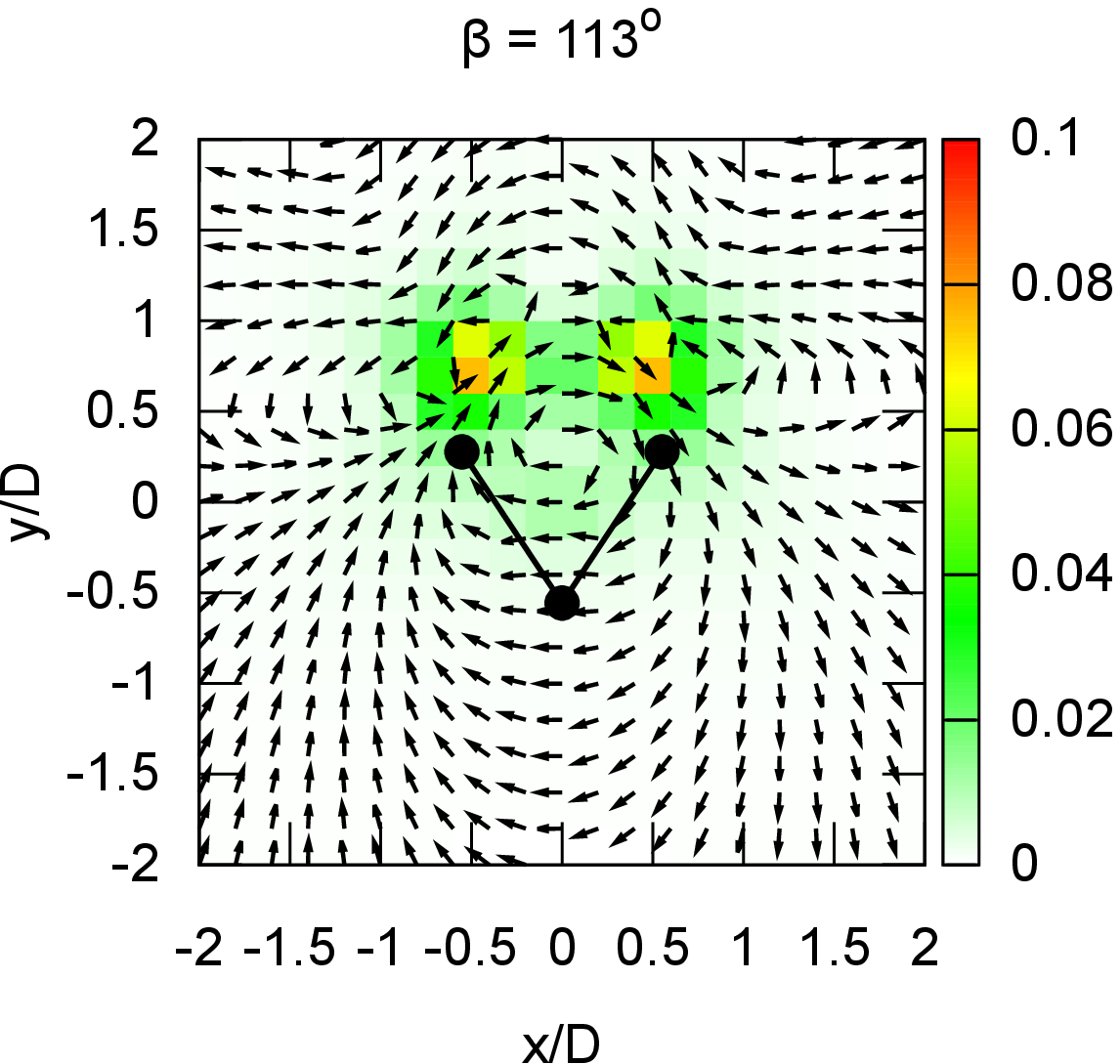}  \includegraphics[width=0.4\textwidth]{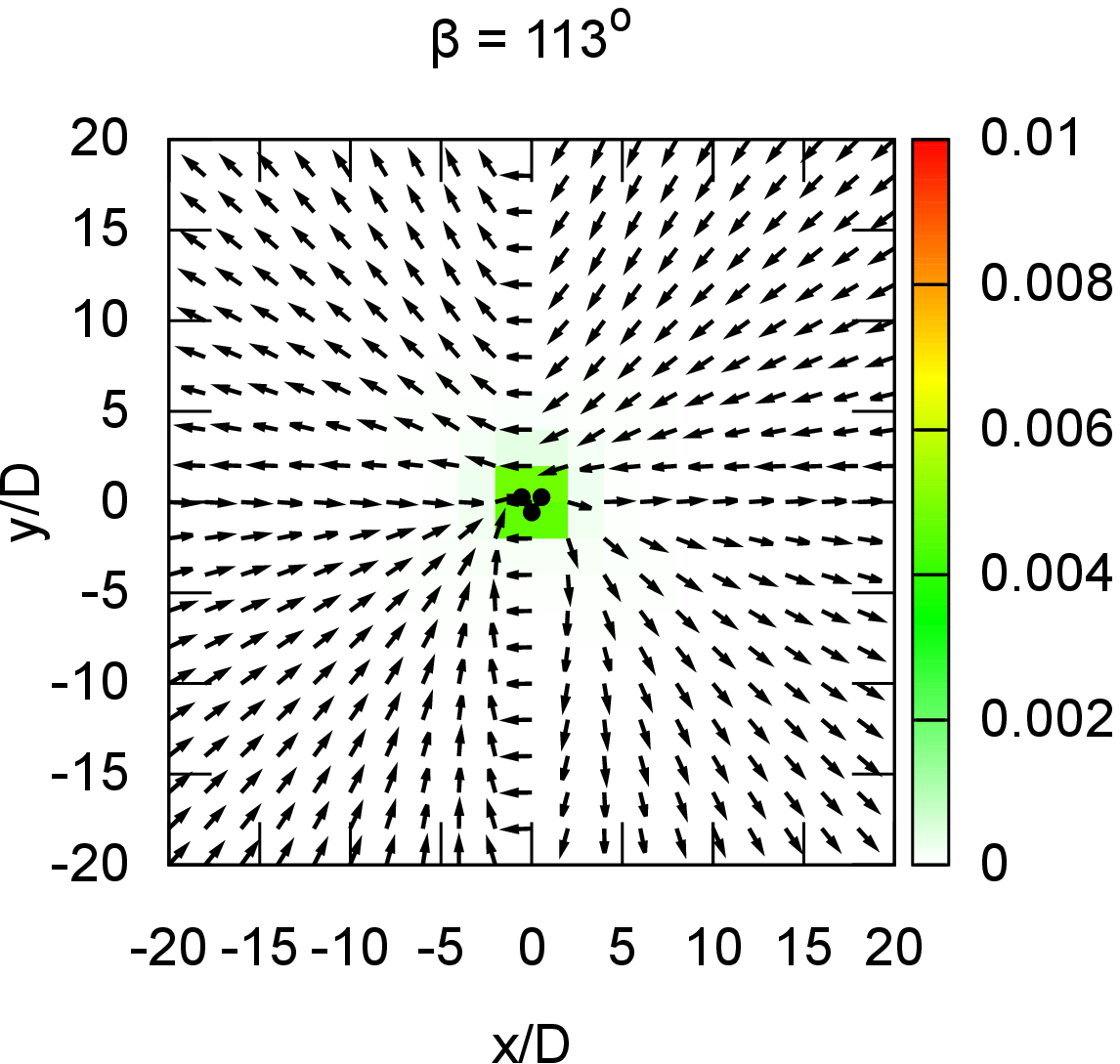}\\
\includegraphics[width=0.4\textwidth]{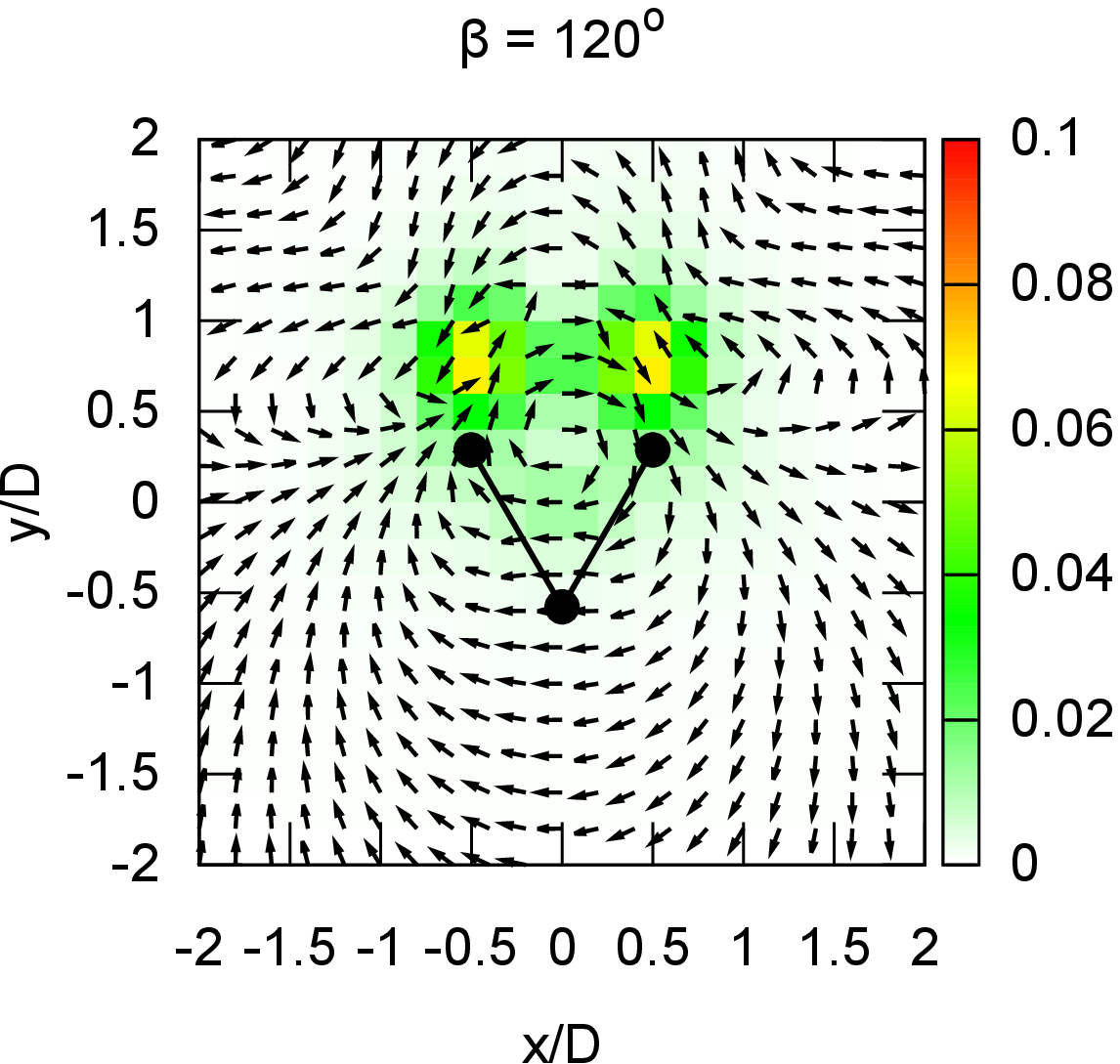}  \includegraphics[width=0.4\textwidth]{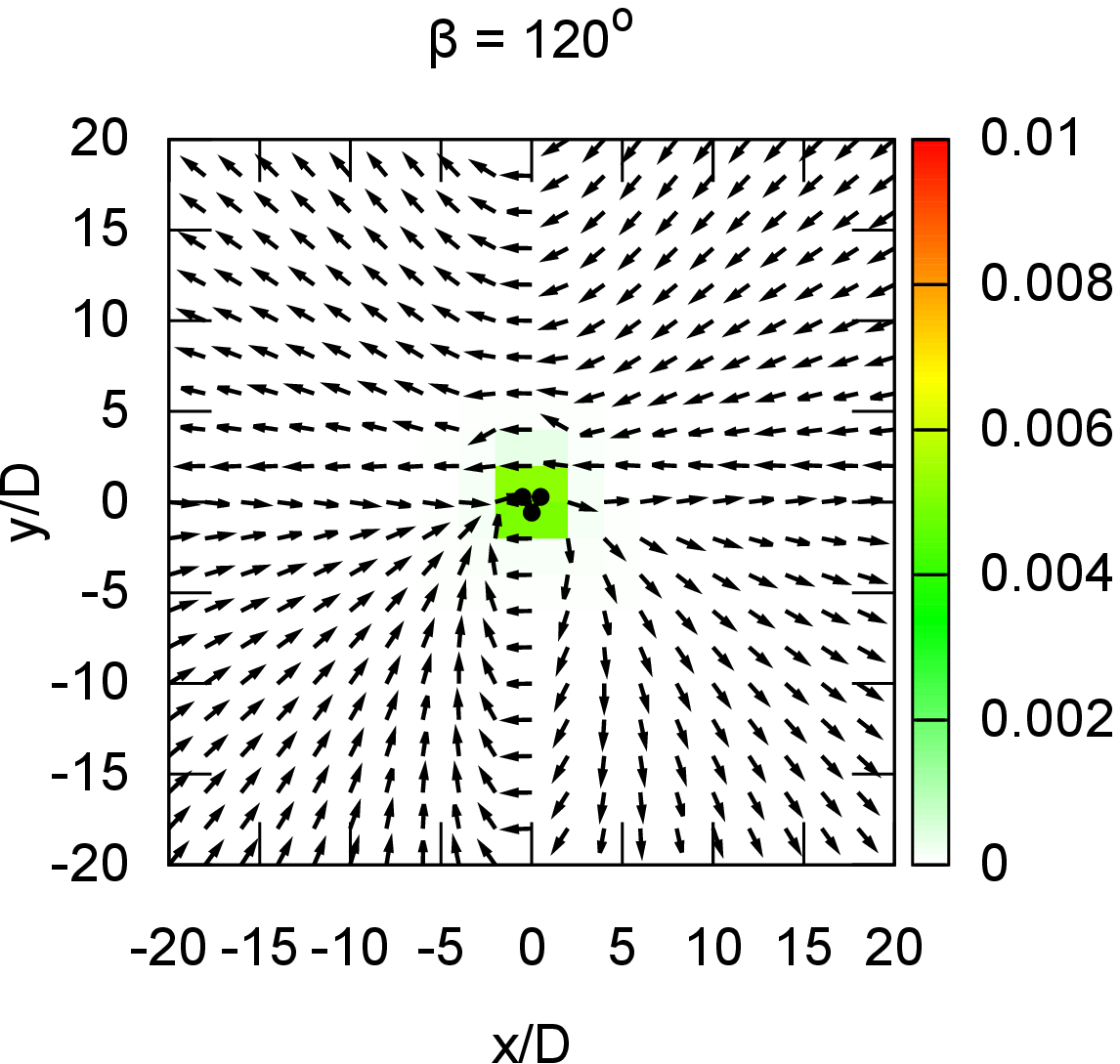}\\
\includegraphics[width=0.4\textwidth]{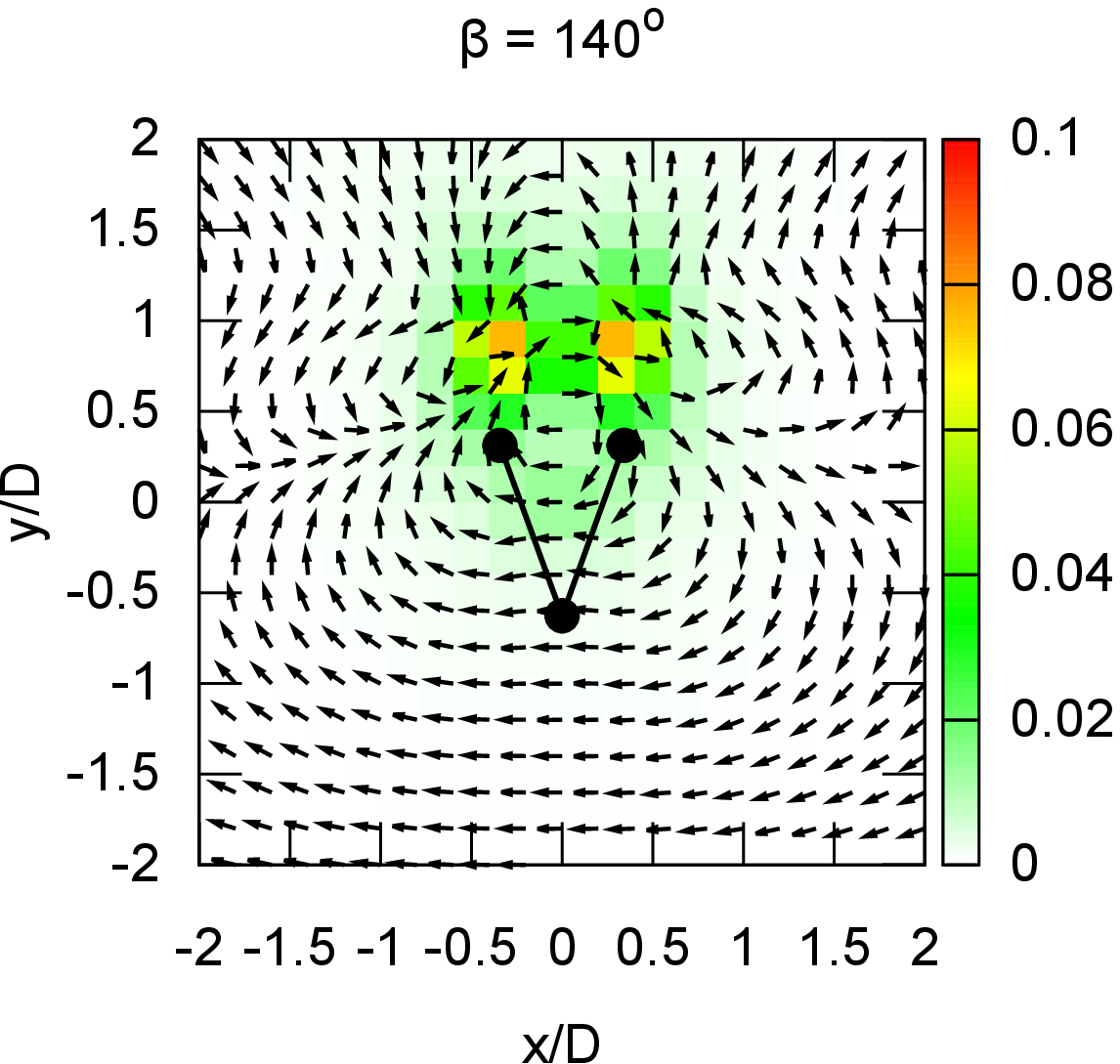}  \includegraphics[width=0.4\textwidth]{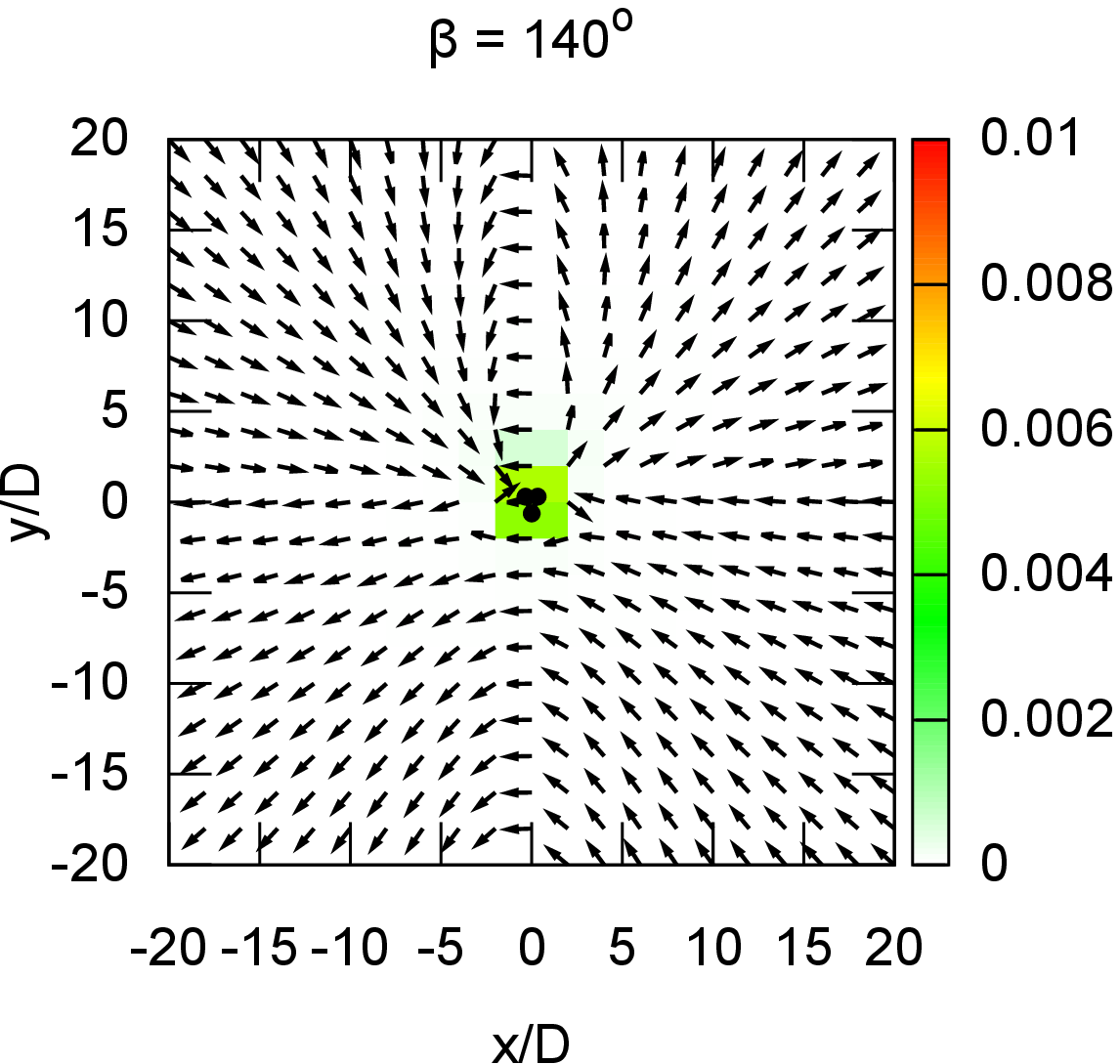}\\
\caption{Average velocity fields near (left) and far (right) from the swimmer for $\beta=113^\circ$, $120^\circ$, and $140^\circ$.  Arrows indicate the 
direction of the velocity field, $\mathbf v_i/|\mathbf v_i|$,while the colour scale indicates its normalised magnitude, $P|\mathbf v_i|/\epsilon$.   \label{fig:velocityfields2}}
\end{figure*}

Figs.~\ref{fig:velocityfields1} and~\ref{fig:velocityfields2}  depict the direction (arrows) and magnitude (colours) 
of $\bar{\mathbf{v}}_i $ for $\beta=\{0^\circ,53^\circ,107^\circ\}$ and $\beta=\{113^\circ,120^\circ,140^\circ\}$, respectively.   We plot the velocity field at two different length scales:  on the left we 
show the range $-2D < x,y<2D$, and on the right we show the range  $-20D < x,y<20D$.    At small length scales the 
strength of the velocity field is higher close to beads 1 and 3, and is weakly dependent on $\beta$.  The flow pattern resembles that of the linear
swimmer for small $\beta$, for $\beta>107^\circ$  the swimmers exhibit a recirculation pattern, up to distances comparable to the swimmers body length.   
This behaviour, however, is rapidly lost at larger length scales, as shown in the left panels of  Figs.~\ref{fig:velocityfields1} and~\ref{fig:velocityfields2}. 

\begin{figure*}
\centering
\includegraphics[width=0.45\textwidth]{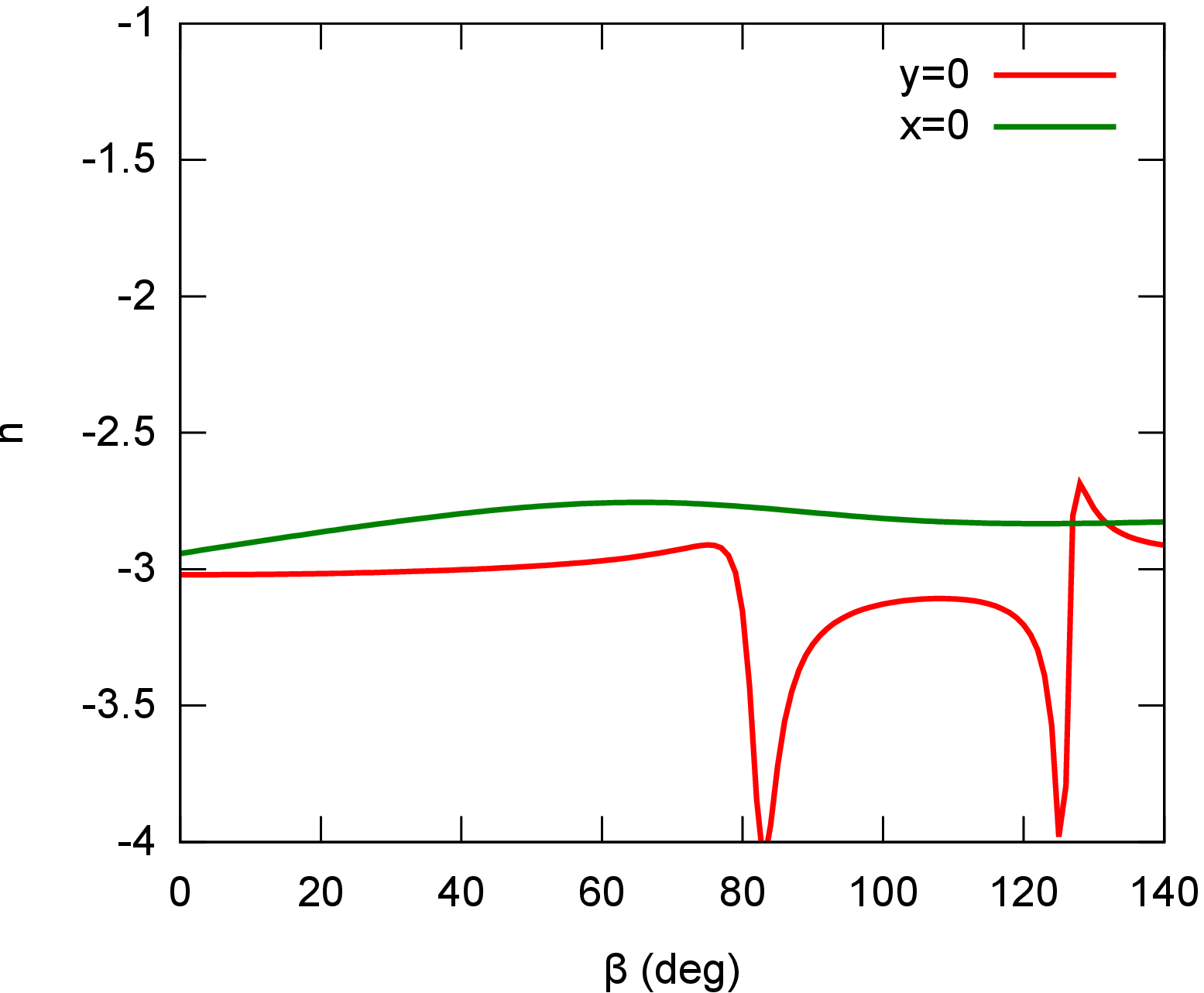} \includegraphics[width=0.45\textwidth]{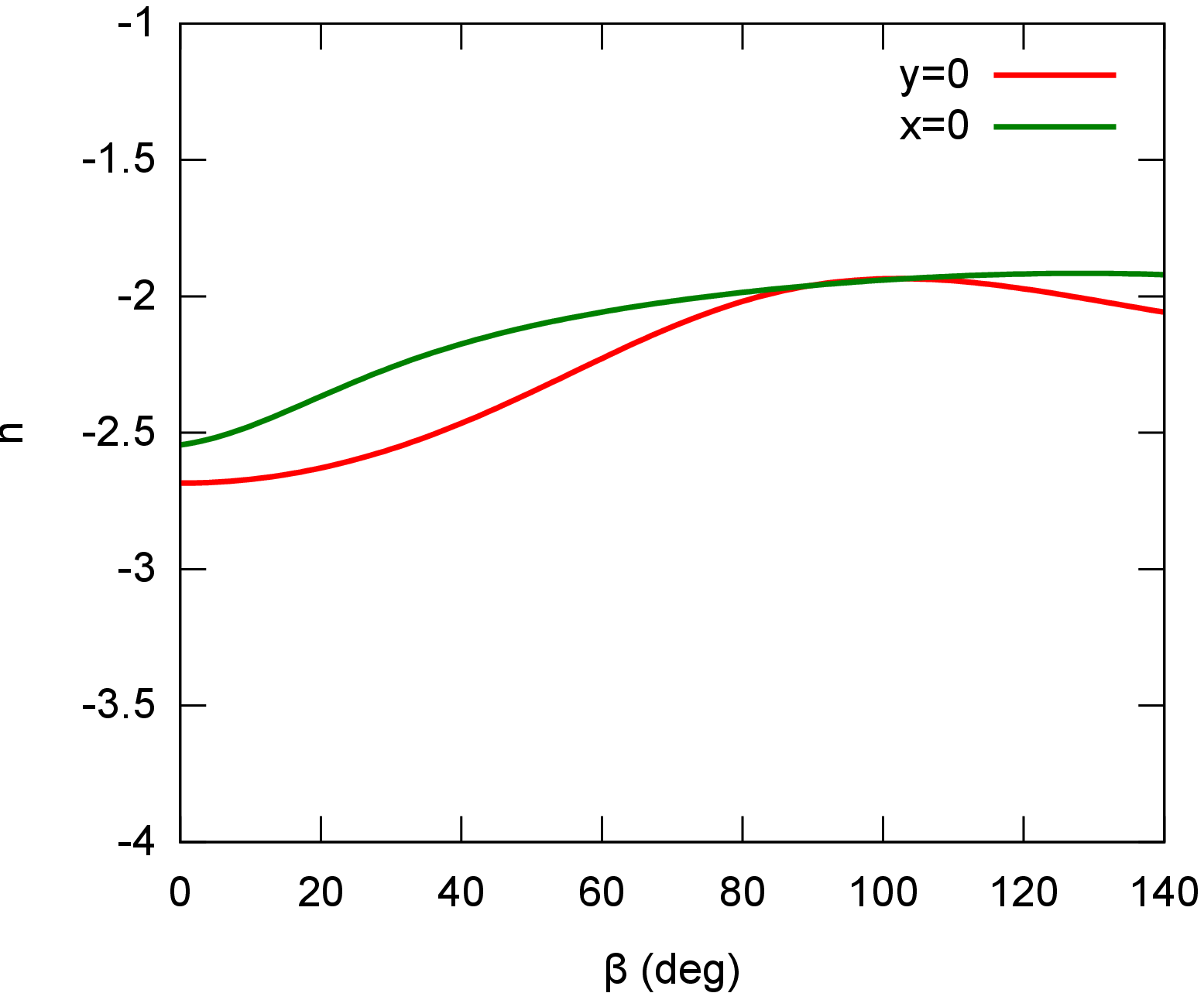}\\
(a)\hspace{6cm}(b)
\caption{Apparent exponent governing the decay of the magnitude of the velocity field along the $x$- and $y-$axes with distance, as a function of the swimmer angle $\beta$.  (a) Symmetric swimmer
with identical maximum arm extension, $D$.  (b) Asymmetric swimmer with maximum arm extensions $0.8D$ and $D$, for $l_1$ and $l_2$. \label{fig:exponents}}
\end{figure*}

We consider now the far-field behaviour. Quadrupolar decays (a velocity field which scales with distance as $r^{-3}$) arise generally for swimmers whose stroke is invariant under a combined time-reversal and parity transformation~\cite{scattering1}. This invariance must be reflected in the velocity field -- resulting in odd-power decays. For our swimmer the symmetry holds when the extension and deformation of both legs are identical.  While the flow patterns for $\beta>0$ are in general reminiscent of a stresslet velocity field, a closer inspection of the magnitude of the
velocity shows that it does indeed have a power-law decrease with distance governed by an exponent $n=-3$.
Fig.~\ref{fig:exponents}(a) plots the apparent decay exponent, $n$, measured at a large distance from the swimmer, as a function of $\beta$.  We consider the decay 
along the the $x$- and $y$-axes, and define $n$ as
\begin{equation}
n (y=0) = \frac{\partial \ln {|\mathbf v}_i|} {\partial \ln x} \mid_{y=0},
\end{equation}
and
\begin{equation}
n (x=0)= \frac{\partial \ln {|\mathbf v}_i|} {\partial \ln y} \mid_{x=0}.
\end{equation}
Since the exponent is measured at a finite distance from the swimmer, its value is in general non-integer.  However, for sufficiently long measuring distances 
we expect to recover a single integer value for $n$. 
Along the $y$-axis the apparent exponent is always closer to $n=-3$, indicating that the quadrupolar term in the velocity field dominates at long distances as expected.
This holds along the $x$-axis as well, except for $\beta\approx\{83^\circ,126^\circ\}$, where higher order terms in the expansion dominate the
decay, as indicated by the larger exponent $n < -4$. This behaviour is interesting, as it suggests that the symmetry of the swimmer at these angles 
suppresses the contribution of the quadrupolar term and gives way to higher order terms in the multipole expansion.

In order to demonstrate the change in the behaviour of the far field velocity as the swimmer loses its invariance under a time-reveral and partity transformation we have measured the apparent exponent for asymmetric swimmers, where $l_1$ and $l_2$ have maximum extensions $0.8D$ and $D$, while all other parameters are kept as before. The results, presented in
Fig.~\ref{fig:exponents}(b), show apparent exponents along the $x$- and $y$-axes, which are consistent with a dipolar decay ($n=-2$).

\section{Conclusions}
\label{sec:Conclusions}
\begin{figure*}
\centering
\includegraphics[height=0.3\textwidth]{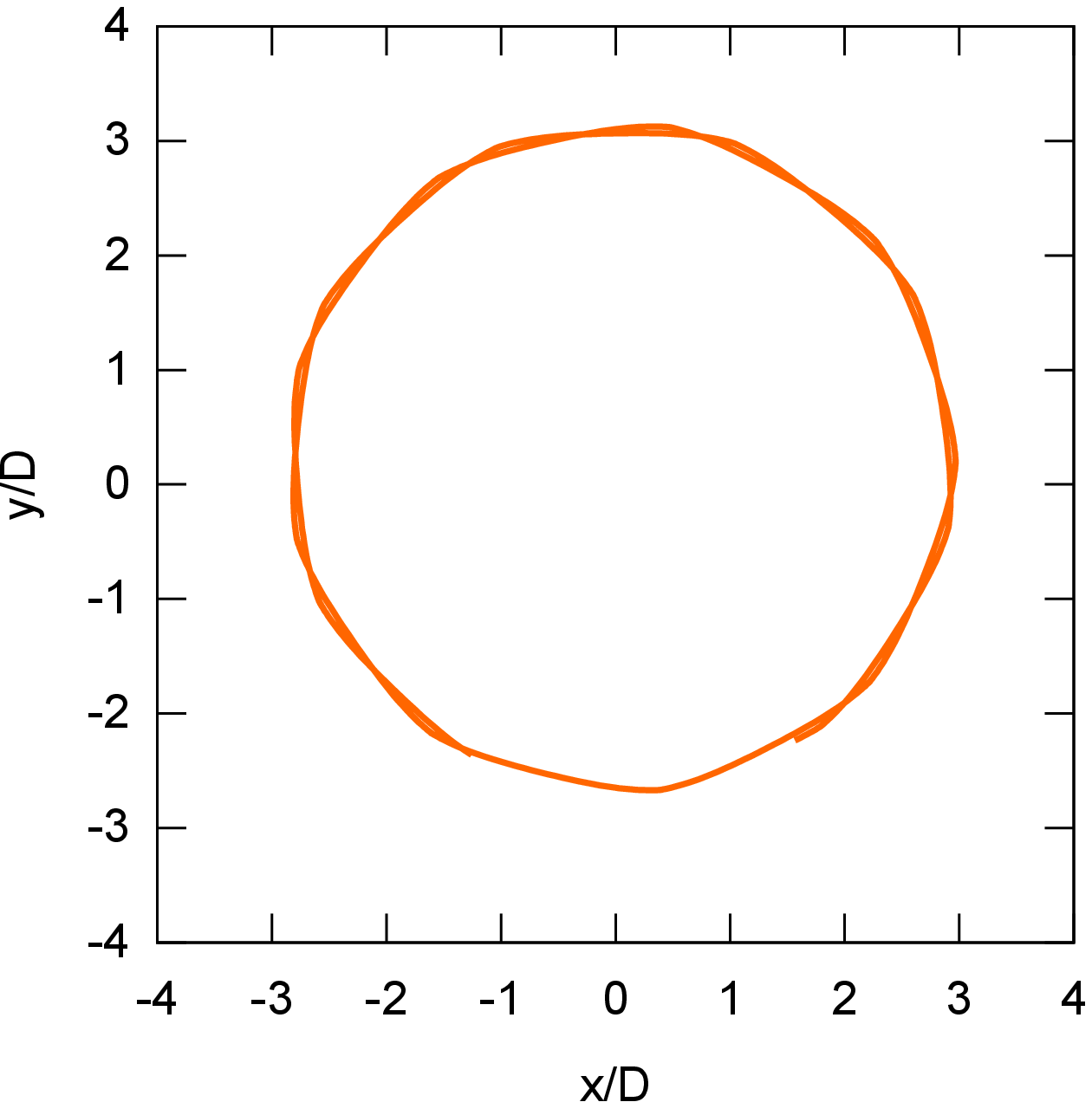} \includegraphics[height=0.3\textwidth]{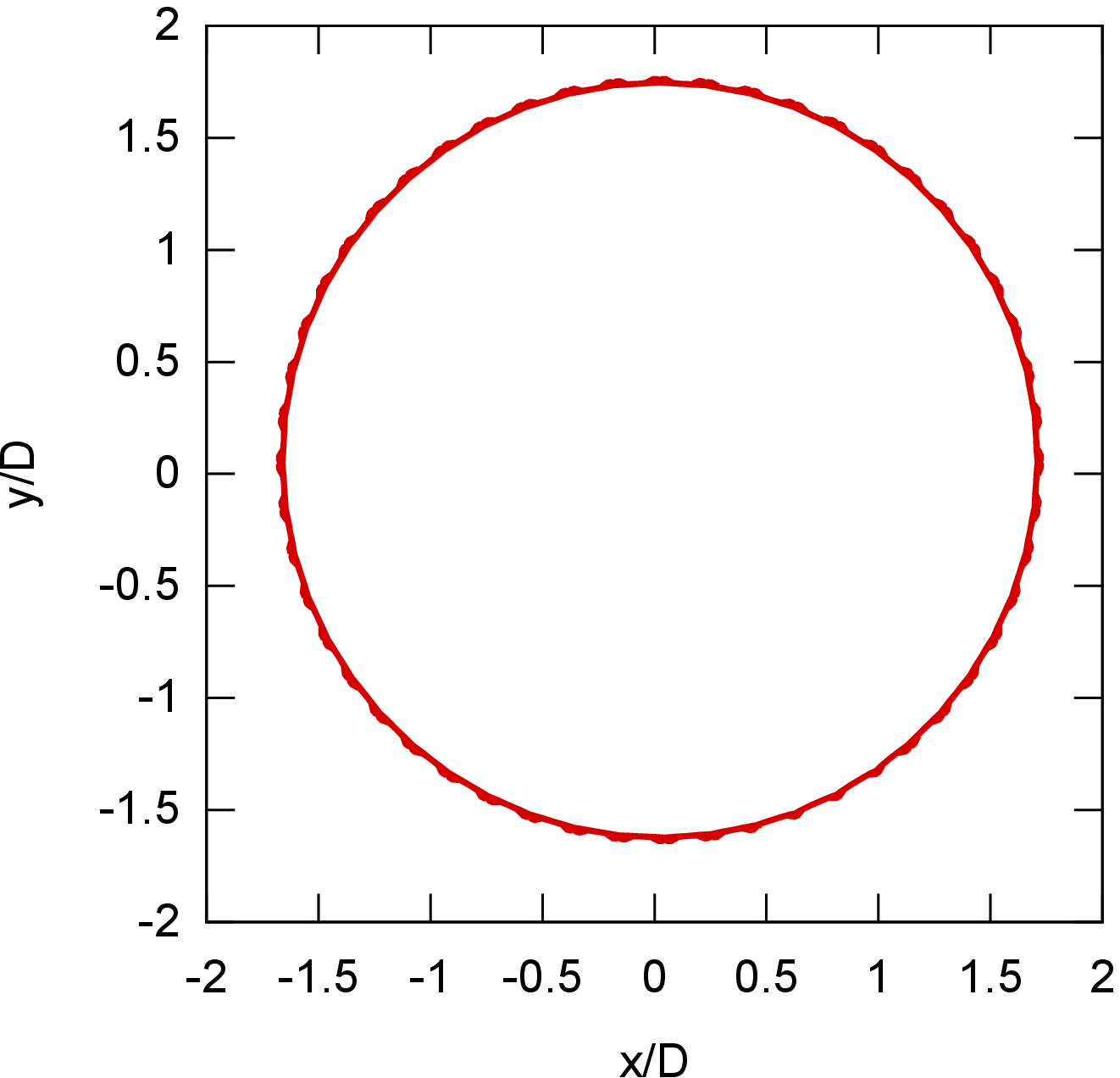} \includegraphics[height=0.295\textwidth]{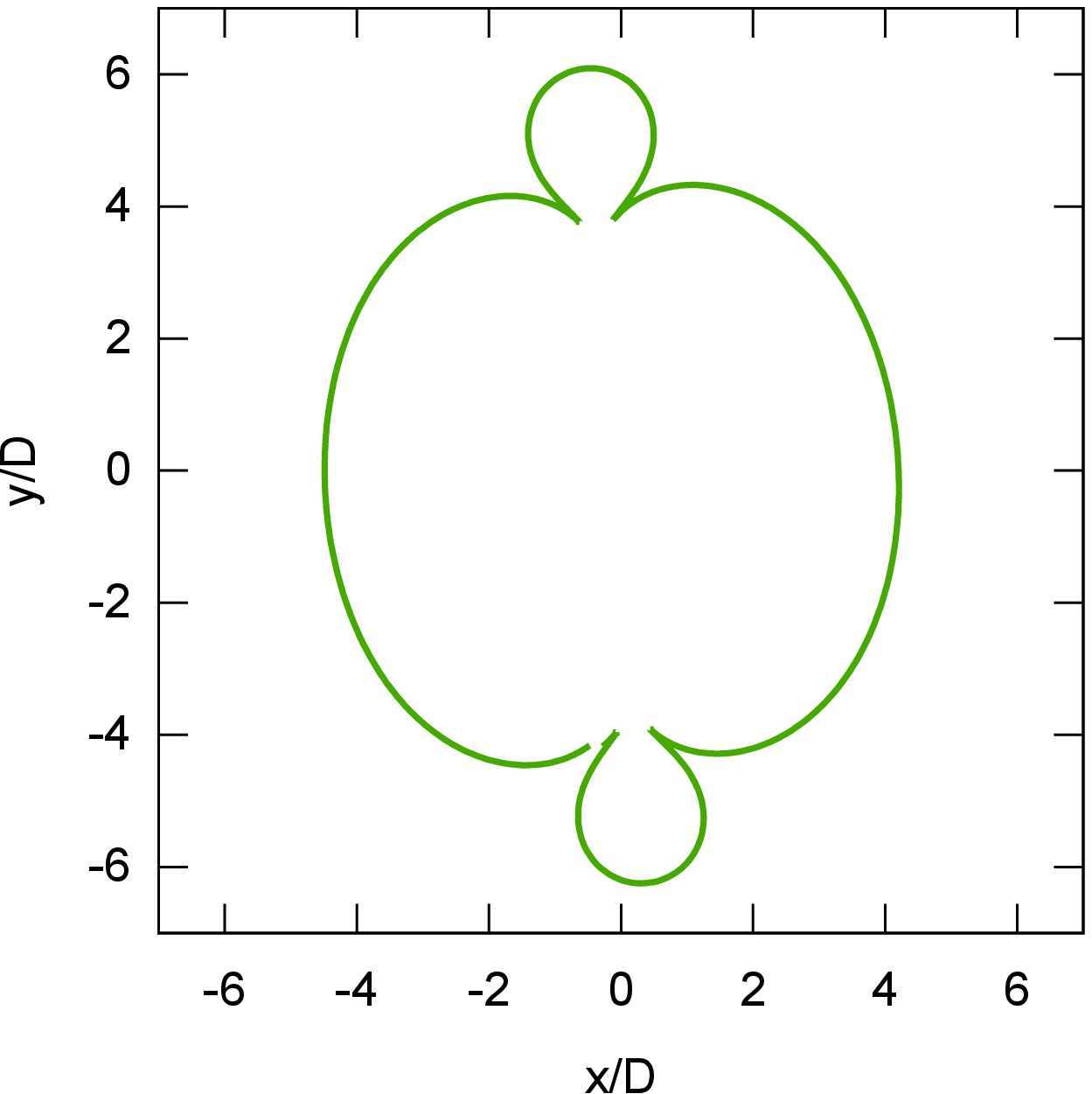}\\
(a)\hspace{5cm}(b)\hspace{5cm}(c)
\caption{Example of swimming pattern where the { swimmer angle} of the swimmer varies according to $\beta(t)=\beta_0+\hat\beta\sin(\Omega t)$, 
with (a) $\beta_0=53^\circ$, $\hat \beta =\beta_0/2$ and $\Omega=2\pi/1000P$, (b) $\beta_0=80^\circ$, $\hat \beta =\beta_0/2$, and $\Omega=2\pi/1000P$ and (c) $\beta_0=80^\circ$, $\hat \beta =4\beta_0/5$ and $\Omega=2\pi/50000P$ . \label{fig:Spiro}}
\end{figure*}

In this paper we have proposed a simple
three-sphere model for a circle swimmer which is a natural generalisation of the linear
Najafi-Golestanian swimmer \cite{Najafi_2004}. The spheres are placed on a triangle
such that two subsequent strokes are performed at an angle.
We find that the radius and the sense of rotation of the swimmer trajectory
depend delicately on the separation angle between the rods joining the beads.
The velocity field produced by this simple circle swimmer
exhibits a characteristic inverse-power decay at large distances.  For swimming strokes invariant 
under a combined time-reversal and parity transformation we recover the expected quadrupolar decay for the velocity field, 
except for a narrow range of separation angles where a stronger decay is observed.  
As expected, for asymmetric swimming strokes, which do not posses the time-reversal and parity symmetry, we recover 
a decay consistent with a dipolar velocity field.  

The model can serve as a simple reference to help understand more complex situations such as a circle swimmer in a confining geometry or the 
collective properties of many circle swimmers.  A further extension of the model is to swimmers whose angle varies in time~\cite{Dreyfus-EPJB-2005,Earl-JChemPhys-2007}. 
Here we expect that more complicated modes of motion arise, depending on the interplay between the timescale of the link variation and the
timescale of the { swimmer angle} variation.  Since the latter essentially controls the radius of the swimming trajectory, a slow variation of the { swimmer
angle} relative to the links result in more complex swimming patterns, as shown in~Fig.~\ref{fig:Spiro}, where the trajectories have a varying curvature leading
to meandering motion.  { Any arbitrary two-dimensional trajectory can be generated by a suitable choice of $\beta(t)$.
Hence $\beta(t)$ can be used as a `steering wheel' to navigate at will. Similar ideas, related to the controllability of the swimmer trajectories, 
have been recently explored by Alouges {\it et al.}~\cite{Alouges}.}   
The motion of circle swimmers under shear ({\it e.g.}, linear shear or Poiseuille flow) can lead to new trajectories such as cycloids. Other external driving that could alter trajectories include a gravitational field or 
magnetic or electric fields~\cite{Lowen_JPCM}. One is just beginning to understand the topologies
of these trajectories for simple noise-free circle swimmers~\cite{Hagen_2011_PRE}.  

The trajectories of the circle swimmers studied in this paper could also be verified experimentally, {\it e.g.}, by controlling the trajectories
of immersed microbeads using optical traps, as performed by Leoni {\it et al.}~\cite{Leoni-SoftMatter-2008} for the linear three-bead swimmer, 
and there is no obstacle in principle to do this for our circle swimmer as well. 

Finally, putting together more than three spherical beads could lead to more complicated motion, in three dimensions,
than the simple circular trajectories discussed in this paper. This results from the intricate translation-rotation
coupling for biaxial particles. It was shown that the simple Brownian circle swimmer
in three spatial dimensions possesses a wealth of different trajectories with the circular helix
being the simplest one \cite{Wittkowski_2012}. It would be interesting to generalise our model
to four spheres which are not in a common plane in order to access these complicated types of motion.

\section{Acknowledgement}

RL acknowledges funding from Marie Curie Actions (FP7-PEOPLE-IEF-2010 no. 273406), JMY from the ERC Advanced Grant (MiCE) and HL from the DFG within the SFB TR6 (project D3).

%
% BibTeX users please use
% \bibliographystyle{}
% \bibliography{}
%
% Non-BibTeX users please use

\end{document}